\pdfoutput=1
\documentclass[12pt]{iopart}
\usepackage{hyperref}
\hypersetup{
pdfauthor=Havemann; Heinz; Struck; Gl\"aser,
pdftitle=Identification of overlapping communities and their hierarchy by locally calculating community-changing resolution levels,
colorlinks=true,
linkcolor=blue,
anchorcolor=black,
citecolor=blue,
urlcolor=blue,
menucolor=black
}

\usepackage[pdftex]{graphicx}

\usepackage{algorithmic}

\usepackage{array}

\usepackage{url}

\begin{document}
%
\title[Identification of overlapping communities]{Identification of overlapping communities\\and their hierarchy by locally calculating community-changing resolution levels}

\author{Frank Havemann$^1$, Michael Heinz$^1$, Alexander Struck$^1$ and Jochen Gl\"aser$^2$}
\address{$^1$ Institut f\"ur Bibliotheks- und Informationswissenschaft, Humboldt-Universit\"at zu Berlin, Unter den Linden 6, 10099 Berlin, Germany, \ead{frank.havemann (at) ibi (dot) hu-berlin (dot) de}}
\address{$^2$ Zentrum f\"ur Technologie und Gesellschaft, Technische Universit\"at Berlin, Hardenbergstr. 36A, 10623 Berlin, Germany}

\begin{abstract}
We propose a new local, deterministic and parameter-free algorithm that detects fuzzy and crisp overlapping communities in a weighted network and simultaneously reveals their hierarchy. Using a local fitness function, the algorithm greedily expands natural communities of seeds until the whole graph is covered. The hierarchy of communities is obtained analytically by calculating resolution levels at which communities grow rather than numerically by testing different resolution levels. This analytic procedure is not only more exact than its numerical alternatives such as LFM and GCE but also much faster. Critical resolution levels can be identified by searching for intervals in which large changes of the resolution do not lead to growth of communities. We tested our algorithm on benchmark graphs and on a network of 492 papers in information science. Combined with a specific post-processing, the algorithm gives much more precise results on LFR benchmarks with high overlap compared to other algorithms and performs very similar to GCE.
\end{abstract}
\submitto{J.\,Stat.\,Mech.}

\maketitle

\tableofcontents

\section{Introduction}
Many real-world networks consist of substructures that overlap because nodes are members of more than one substructure. Networks of scientific papers are a case in point. Thematic structures such as topics, approaches, or methods are not disjunct. It is the rule rather than the exception that a paper addresses more than one topic, refers to more than one conceptual approach, or reports the use of more than one method. 

Hard clustering is inadequate for the investigation of real-world networks with such overlapping substructures. Instead, methods are required that allow nodes to be members of more than one community (or module) in the network. Beginning with the well-known work by Palla, Derényi, Farkas, and Vicsek in 2005 \cite{palla2005uoc}, a number of approaches to the detection of overlapping communities in graphs have been implemented and tested. One approach starts from hard clusters obtained by any clustering method and fractionally assigns the nodes at the borders between clusters to these clusters~\cite{baumes2005efficient,wang2009adjusting}. Another approach is based on a hard clustering of links into disjoint modules, which makes nodes members of all modules their links belong to \cite{ahn2009link,evans2009edge}. A third approach, which inspired our work, constructs \textit{natural communities} of all nodes, which can overlap each other \cite{lancichinetti2009detecting}. A natural community of a node is a module that was grown from it by a greedy algorithm that maximises a resolution-dependent local fitness.  

In our search for methods that model scientific specialties as networks of journal papers and enable the identification of thematic structures in those networks, we use the approach that identifies natural communities because it is well suited to our problem. The construction of natural communities of nodes can be interpreted as the construction of a thematic environment from the `scientific perspective' of the seed paper. This idea is not only attractive from a conceptual point of view---the borders of topics are explored by a local algorithm i.e.\ independently from papers located far away from the seed paper---but also for services leading users of bibliographic databases from one relevant paper to thematically similar ones. Furthermore, this approach identifies not only overlapping communities but also a hierarchical structure of a graph if there is any. Since we assume that thematic structures are of varying scope and that some of the smaller themes might be completely contained in larger ones, an algorithm that detects both overlaps and hierarchies is essential.

However, while the basic approach of constructing overlapping substructures as natural communities of nodes is very attractive, the algorithms using this idea that have been proposed so far are less then optimal. The original algorithm proposed by Lancichinetti, Fortunato and Kertesz (LFK)~\cite{lancichinetti2009detecting} partially violates the principle of locality and is based on a numerical procedure. A variation proposed by Lee, Reid, McDaid, and Hurley \cite{lee2010detecting}---Greedy Clique Expansion (GCE)---remedies the locality problem but is still based on a numerical solution. The numerical approach implies that the procedure has to be repeated for each value of the resolution parameter, which is inefficient and provides only estimates of the resolution levels at which the network covers change.

To overcome these problems, we propose an algorithm that is based on the exact calculation of resolution levels at which the membership of natural communities actually changes, and which therefore determines natural communities for all resolution levels in one run. Since this algorithm is based on the \textbf{m}erging of \textbf{o}verlapping \textbf{n}atural \textbf{c}ommunities, we have termed it MONC.\footnote{Some of the results presented were also shown on a poster with the title \textit{A local algorithm to get overlapping communities at all resolution levels in one run} at ASONAM conference, Odense, Denmark, August 2010 \cite{havemann2010local}.}

\section{Algorithms for the detection of natural communities}

\subsection{Numerical approaches: LFK, LFM, and GCE}
The essence of the LFK algorithm is that independently constructed natural communities of nodes can overlap. In accordance with the locality of their approach Lancichinetti, Fortunato, and Kertesz evaluate the fitness of modules of nodes with a function that uses only local information. The fitness function is based on the premise that communities are dense subgraphs with only sparse links between them. It is defined as the ratio of the sum of internal degrees to the sum of all degrees of nodes in a module $G$. The denominator is taken to the power of $\alpha$, the resolution parameter: 

\begin{equation} \label{LFK-fitness} 
f(G, \alpha) = \frac{k_{in}(G)}{(k_{in}(G) + k_{out}(G))^\alpha}.
\end{equation} 

For each node a \textit{natural community} $G$ is constructed by including the neighbour that produces the highest fitness gain. Then the fitness gain of each node in $G$ is recalculated, and nodes with negative fitness are removed from $G$. The community is complete if including any neighbour brings no fitness gain. 
The authors conclude \cite[p.\ 6]{lancichinetti2009detecting}: ``By varying the resolution parameter one explores the whole hierarchy of covers of the graph, from the entire network down to the single nodes, leading to the most complete information on the community structure of the network.'' 

Since the LFK algorithm constructs natural communities of all nodes of a graph and has to be repeated for each value of the resolution parameter within the interval of interest, applying it to larger networks is time-consuming. Acknowledging this, the authors proposed several ways in which their algorithm could be optimised. For example, they proposed to use communities found at one level of resolution as starting points for the next lower level because at lower resolution a community cannot be smaller than at a higher level~\cite[p.\ 7]{lancichinetti2009detecting}. Another proposed solution---the only one they implemented and tested---starts from a random node and after the construction of its community randomly selects another seed node outside this community until the whole graph is covered. This random version of the LFK algorithm is usually denoted by \textit{LFM} (for local fitness maximisation).

The LFK and the LFM algorithms partly violate the principle of locality in two respects. First, their fitness function does not allow nodes to stay singles because for any $\alpha$ the module fitness of a single is always zero and the module fitness of two neighbours is always larger then zero. Second, the algorithms recheck the fitness contribution of all community nodes after a new node has been added and exclude nodes if their removal increases the fitness. This can even lead to the exclusion of a seed node from its own natural community. In our networks of papers, removing nodes that reduce the fitness of a grown community would be equivalent to shifting from the individual thematic perspective of the seed paper to a collective perspective of all papers in the community. 

Lee \etal \cite{lee2010detecting} implemented the LFK algorithm without exclusion mechanism. Their GCE algorithm also differs from LFK in that it starts from maximum cliques of nodes rather than single nodes. GCE includes a threshold-dependent procedure that removes communities from further analysis which are near-duplicates of communities already found (s. \textit{Experiments} section, p.\ \pageref{delta}).\footnote{In the \textit{Further Work} section of reference \cite[p.\ 9]{lee2010detecting} the authors mention that they are working on a version of their algorithm which---like MONC---expands all seeds in parallel. However, it is not yet known whether this approach is based on  numerical or exact solutions for the resolution parameter.)}

LFK, LFM, and GCE share a numerical approach, i.e.\ they can detect the hierarchical module structure of a graph only if the resolution parameter $\alpha$ is changed in small steps and the community detection procedure is repeated for each level of $\alpha$. This turns out to be an unnecessary  operating expense and loss of precision, as we will demonstrate by presenting an algorithm that is based on the calculation of exact solutions of the necessary (and only the necessary) levels of the resolution parameter.

\subsection{An algorithm based on exact solutions for community-changing resolution levels: MONC}
We assume that each node is its own natural community $G$ at infinite resolution. The next vertex $V$ from the neighbourhood of $G$ included to $G$ is the one that increases the fitness of $G$ at the largest value of resolution denoted by $\alpha_\mathrm{incl}(G, V)$. 
In pseudo code the growth of a natural community $G$ can be described as follows ($N(G)$ denotes the neighbourhood of $G$): 

 \begin{algorithmic}[1]
 \WHILE {$N(G)$ is not empty} \FOR {each node $V$ in $N(G)$} 
 \STATE calculate $\alpha_\mathrm{incl}(G,  V)$ 
 \ENDFOR \STATE include the node with maximum $\alpha_\mathrm{incl}$ into $G$ 
 \ENDWHILE \end{algorithmic} 

If two nodes have equal $\alpha_\mathrm{incl}$ MONC should include both. We did not implement this for the experiments described below, which is why the second node is included at a marginally lower $\alpha_\mathrm{incl}$. The pseudo code shows that MONC does not enable the removal of nodes from a natural community. As we have argued above, this possibility contradicts the principle of locality. The complete MONC pseudo code can be found in the \textit{Supplementary information} section (p. \pageref{pseudo-code}).

Since nodes cannot remain singles if the fitness function as defined by Lancichinetti \etal  \cite{lancichinetti2009detecting} is used, we slightly changed $f(G,  \alpha)$ (cf.\ eq.\ \ref{LFK-fitness}) by adding 1 to the numerator:\footnote{We could also avoid this artefact of LFK's fitness function by adding self-links to all nodes i.e.\ we would assume that a node is most similar to itself.}
 \begin{equation} 
 f(G, \alpha) = \frac{k_\mathrm{in}(G) + 1}{(k_\mathrm{in}(G) + k_\mathrm{out} (G))^\alpha}.
 \end{equation} 
From this definition we can derive a formula for calculating the maximum value of resolution $\alpha_\mathrm{incl}(G, V)$, where a node $V$ does not diminish the fitness of a module $G$ when included in it by demanding that for $\alpha < \alpha_\mathrm{incl}(G, V)$ we have $f(G \cup V, \alpha) > f(G, \alpha)$ (cf. \textit{Supplementary information}): 
 \begin{equation} 
 \alpha_\mathrm{incl}(G, V)  =  \frac{\log( k_\mathrm{in}(G \cup V) + 1 ) - \log (k_\mathrm{in}(G) + 1)}  {\log k_\mathrm{tot}(G \cup V) - \log k_\mathrm{tot}(G) },
\label{alpha_incl}
 \end{equation} 
where $k_\mathrm{tot}=k_\mathrm{in}+k_\mathrm{out}$ denotes the sum of the degrees of all nodes of a module. 

\subsubsection*{Optimisation of MONC:}
We can calculate $k_\mathrm{in}(G \cup V)$ from $k_\mathrm{in}(G)$ and $k_\mathrm{tot}(G \cup V)$ from $k_\mathrm{tot}(G)$ i.e.\ the current values of the module from the preceding ones (which saves computing time). For this we define the interaction of a module and a node as 
 \begin{equation} 
 k_\mathrm{inter} (G, V) = \sum_{i \in G}A_{Vi},
 \end{equation} 
where $A$ denotes the adjacency matrix of the undirected (and in general) weighted graph, and calculate the degree of a node or its weight as the sum of the weights of its edges to all $n$ nodes of the graph: 
 \begin{equation} 
 A_{V+} = \sum_{i=1}^nA_{Vi}. 
 \end{equation} 
The weight of edges of internal nodes $k_\mathrm{in}$ is increased by $2\cdot k_\mathrm{inter}$ because both directions have to be taken into account: 
 \begin{equation} 
 k_\mathrm{in}(G \cup V) = k_\mathrm{in}(G) + 2\cdot k_\mathrm{inter} (G, V).   
 \end{equation} 
The total of all weights is increased by the weights of the edges of the new node:  
 \begin{equation} 
 k_\mathrm{tot}(G \cup V) = k_\mathrm{tot}(G) + A_{V+}.
 \end{equation} 
We first include the neighbour $V$ of each node that improves the community's fitness at highest resolution. Then we continue with the new neighbourhood of $G\cup V$ until all nodes are included in the natural community. After each step we compare the current communities of all nodes to find duplicates. Thus we can reduce the number of communities treated by the inclusion algorithm and save further computing time. We merge overlapping natural communities of nodes. 

\subsubsection*{Cliques as seeds:}
If local densities of a graph strongly vary and the seed node is located in a high density region, MONC leaves this region when including the second node. It searches for nodes with low degree first because these nodes only moderately increase the number of links leaving the community and thus often provide the earliest increase in fitness. We surmise that the LFK algorithm `repairs' this unwanted behaviour by allowing the exclusion of nodes with negative fitness. Since we suppressed the exclusion of nodes, we solved this problem by starting from cliques (i.e.\ totally linked subgraphs) instead of single nodes. 

As noted above,  Lee \etal \cite{lee2010detecting}, whose GCE algorithm is essentially an implementation of LFK without the exclusion mechanism, also found that cliques as seeds gave better results than single nodes. However, while Lee \etal use maximal cliques (i.e.\ cliques which are not subgraphs of other cliques), we optimise clique sizes by choosing the most cohesive subgraph of a clique as seed. This is achieved by excluding the node $V$ that diminishes the module fitness at lowest resolution, i.e.\ has the weakest coupling to the rest of the module $G$. Analogously to $\alpha_\mathrm{incl}$ we calculate $\alpha_\mathrm{excl}$ with 
 \begin{equation} 
 \alpha_\mathrm{excl}(G, V)  =  \frac{\log( k_\mathrm{in}(G) + 1 ) - \log (k_\mathrm{in}(G \setminus V) + 1)}  {\log k_\mathrm{tot}(G) - \log k_\mathrm{tot}(G \setminus V) }.
 \end{equation} 
This procedure is repeated until only two nodes remain in each clique. From the set of shrinking cliques we select the one which is most resistant to further reduction, i.e.\ those with highest $\alpha_\mathrm{excl}$ of the next node to be excluded. After its exclusion the rest of the clique would be less strongly coupled (for details see section \textit{Experiments} and cf. figure \ref{fig_melting-h-clique} and MONC pseudo code 
in \textit{Supplementary information}). Nodes which are not member of any optimal clique remain single seeds. Every other node is assigned to the clique where it has its maximum $\alpha_\mathrm{excl}$.

\subsubsection*{Post-processing:}
After running MONC we need some post-processing to detect definite modules at relevant resolution levels. We discuss post-processing issues together with our experiments, where we can illustrate them with data. The identification of critical resolution levels is described in the section about the karate-club experiment. The detection of crisp and fuzzy modules at relevant levels is described in the section about synthetic benchmark graphs. Crisp modules exist when each individual either belongs or doesn't belong to a community. Modules are fuzzy when an individual's strength of membership in a module may vary between zero and one~\cite[p. 1]{gregory2010fuzzy}.

\subsection{Comparison of the algorithms}
The algorithms LFK, LFM, GCE and MONC look like minor variations in the implementation of the same idea, namely detecting overlapping natural communities of nodes. However, there are two significant differences between the algorithms. First, LFK, LFM and GCE are based on a numerical approach that depends on the resolution parameter, while MONC is based on an exact solution for the resolution parameter, which makes it a parameter-free algorithm. Second, LFK and LFM compromise on the principle of locality, while GCE and MONC are based on strict locality. 

\subsubsection*{Numerical versus exact solutions:} 
To find relevant resolution levels at which the hierarchical module structure of a graph can be detected the three algorithms LFK, LFM and GCE must be processed repeatedly for many resolution levels. Thus, they find only approximate values of the resolution at which graph covers change, and need much time to obtain this result. To remedy the latter problem, Lancichinetti \etal proposed a random version of their algorithm, LFM, which randomly selects nodes for which natural communities are grown. However, this creates a trade-off between the number of nodes in a network and the intervals of the resolution level. The random algorithm is equivalent to the original non-random version only if the number of nodes in a network is small compared to the number of resolution levels for which the algorithm is processed.

Owing to its search for community-changing resolution levels, MONC does not create these problems. Since MONC determines communities for all resolution values in one run, it reveals the hierarchical module structure of the graph not only faster than the original LFK algorithm but also faster than LFM and GCE, particularly for large networks. 

If we follow the reasoning of Lancichinetti \etal \cite[pp. 6--7]{lancichinetti2009detecting} we get $O(n^2 \log n)$ as the worst case complexity of LFM for graphs with $n$ nodes. One factor $n$ is due to the exclusion mechanism and $\log n$ is the order of the number of $\alpha$-levels needed to reveal the hierarchy of the network with $n$ nodes.  If the computing time of the random version (LFM) scales with $n^2 \log n$, the computing time of non-random LFK variants should scale with $n^3 \log n$ and that of GCE with $n^2 \log n$. Computing time of MONC should scale with $n^2$ in the worst case because MONC does not exclude nodes and uncovers the whole hierarchy in one run. Furthermore, MONC saves time due to  merging of communities. We will examine MONC's complexity by applying it to benchmark graphs of different sizes in the future.

A third advantage of MONC besides increased precision and speed is a new approach to the identification of critical resolution parameters. Since all community-changing resolution levels are known, the change of network covers can be treated as a function of the resolution parameter, which in turn allows the identification of `interesting' resolution levels including resolution levels at which sudden changes in the community structure occur and intervals in which all or many communities remain stable. 

The fourth advantage of the MONC algorithm is the opportunity to build whole networks from single nodes. Since it calculates community-changing resolution levels, MONC can start from a single node (e.g. a paper in a publication database), identify a suitable clique, and grow a network from that clique by iteratively processing the neighbourhood of the natural community. This is clearly advantageous for the analysis of publication networks because this analysis needs not begin with arbitrarily delineated sets of papers any more. More generally, whenever the delineation of a network is part of the analysis, MONC offers a parameter-free solution for the identification of the network.

\subsubsection*{Locality:}
 The LFK approach endangers the locality principle that is inherent to the general approach of identifying natural communities in two ways. The first problem is inscribed in the algorithms, which recalculate fitness from the larger perspective of the new natural community and enable the exclusion of nodes (including the original seed node) if their contribution to fitness is negative. MONC and GCE do not allow node exclusion but are confronted by a new problem---the forced digression into sparse regions of the networks in the early growth stage. Both algorithms solve this problem by using cliques rather than single nodes as seeds.

The second problem is inscribed in the fitness function, which forces each node to acquire a second one in the first step. GCE circumvents this problem by using all maximal cliques of a graph as seeds, thereby endangering locality in a different way. MONC also uses seed cliques but keeps the procedure as local as possible by selecting the most coherent subgraphs of cliques as seeds, by letting nodes `choose' seed cliques in which they are most strongly embedded, and by letting nodes that are not members of optimised cliques remain as singletons. As a response to the remaining singletons, we also changed the fitness function as described above.

\section{Data}
To compare our algorithm to LFM of Lancichinetti \etal \cite{lancichinetti2009detecting} we first applied both to the network of social relations of 34 members of the well-known karate club analysed by Zachary \cite{zachary1977information}. As Lancichinetti \etal we used the unweighted version of this network (s. \url{http://networkx.lanl.gov/examples/graph/karate_club.html}).

We also applied LFM and MONC to a network of about 492 papers in the 2008 volumes of six information-science journals with a high proportion of bibliometrics papers (see details in \textit{Supplementary information}). 

In the network of information-science papers, two nodes (papers) are linked if they both have at least one cited source in common. The number of shared sources, which is normalised in order to account for different lengths of reference lists, provides a measure of the thematic similarity of papers. We start from the affiliation matrix $M$ of the bipartite network of 531 papers and their cited sources. To account for different lengths of reference lists we normalise the paper vectors of $M$ to an Euclidean length of one. With this normalisation, the element $a_{ij}$ of matrix $A = MM^\mathrm{T}$ equals Salton's cosine index of bibliographic coupling between paper $i$ and $j$. The symmetric adjacency matrix $A$ describes a weighted undirected network of bibliographically coupled papers. 
The main component of the bibliographic-coupling network of information science 2008 contains 492 papers. Two small components (three and two papers, respectively) and 34 isolated papers are of no interest for our experiments.

To compare the quality of MONC results with those of other algorithms which detect graph covers with overlapping modules we tested MONC for 1100 benchmark graphs. The benchmark graphs---networks of 500 nodes---were constructed by applying the algorithm developed by Lancichinetti, Fortunato and Radicchi (LFR) \cite{lancichinetti2008benchmark}. The proportion of nodes in overlaps vary between 2 and 100 percent. As many empirical networks, these synthetic graphs are characterised by power-law distributions of node degrees and of module sizes. The exponents are parameters of the LFR algorithm; we set $\tau_1=-2$ and $\tau_2=-1$. Other parameters which can be chosen by the user are minimum and maximum module sizes~(here $c_{min}=10$, $c_{max}=50$), average and maximum degree~(here $\langle k \rangle = 16$, $k_{max}=40$), the proportion of links between modules (here $\mu = 0.1$), and the number of modules each node in an overlap can belong to~(here $o_m=2$). We use the same LFR parameters as Steve Gregory  in a recent paper \cite[section IV. B]{gregory2010fuzzy} that compares different algorithms detecting overlapping modules in networks.

\section{Experiments}

\subsection{Karate club}
\begin{figure}[!t]
\centering
\includegraphics[width=3in]{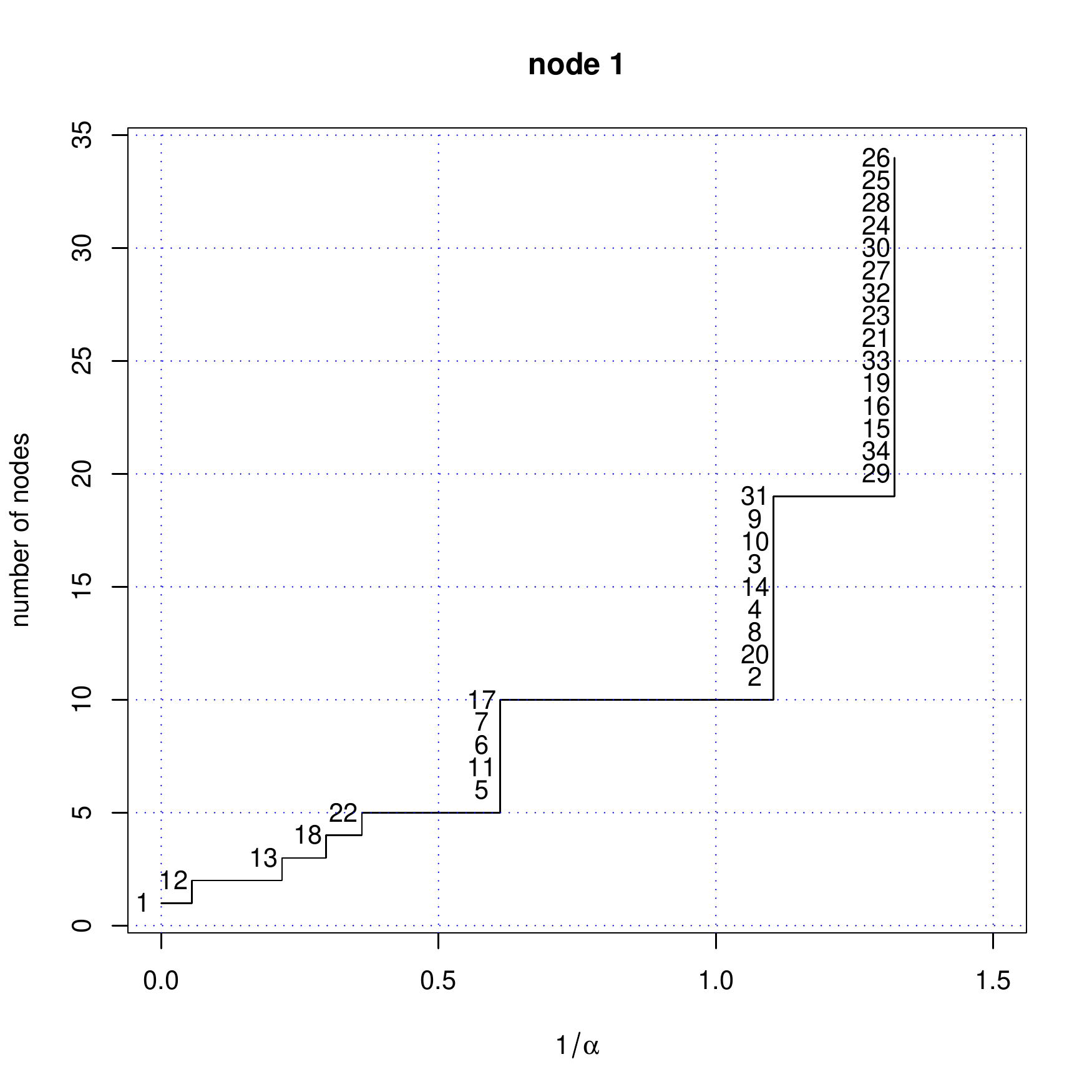}
\includegraphics[width=3.1in]{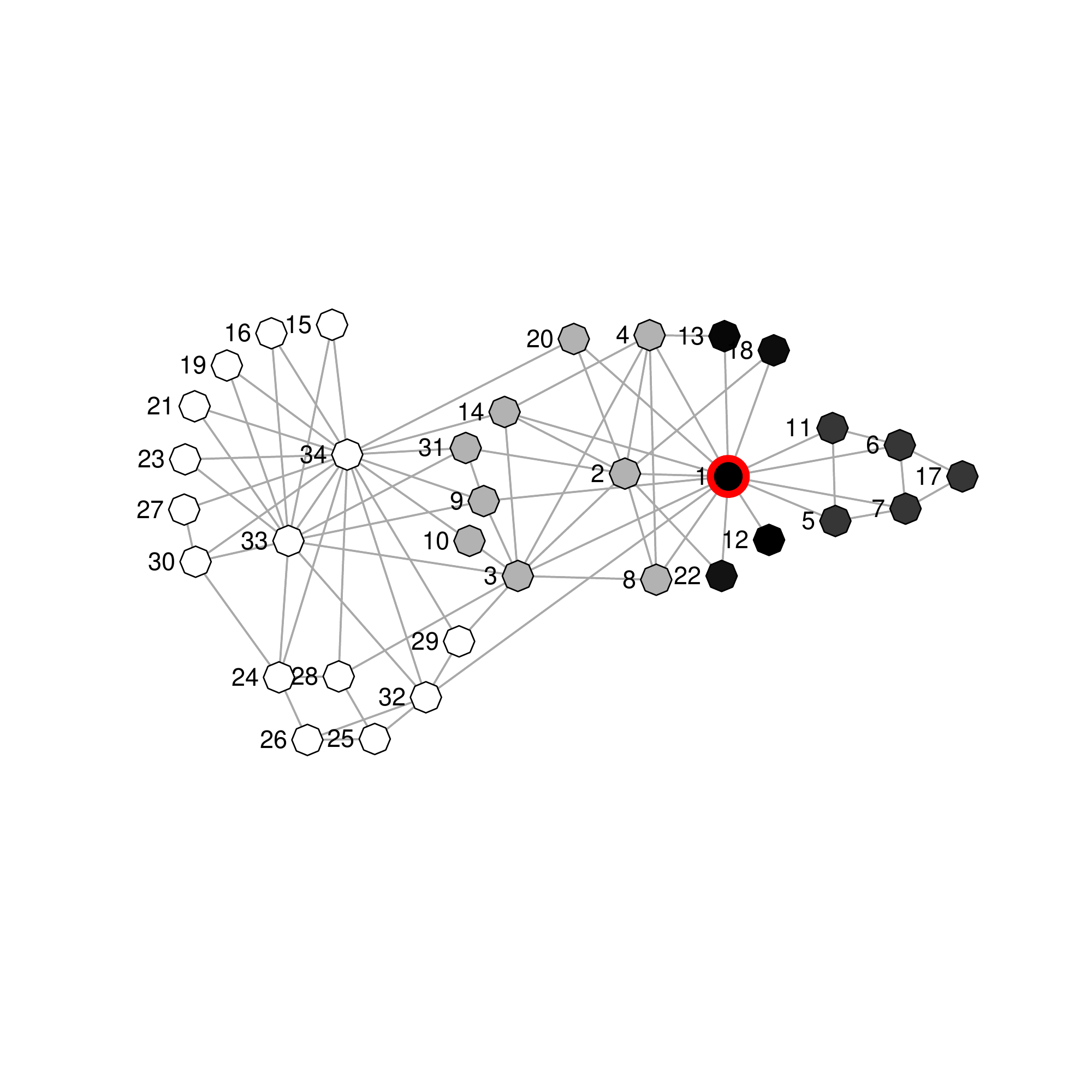}
\caption{Growing natural MONC community of node 1 of karate club}
\label{fig_karate.comm.1}
\end{figure}

\begin{figure}[!t]
\centering
\includegraphics[width=3in]{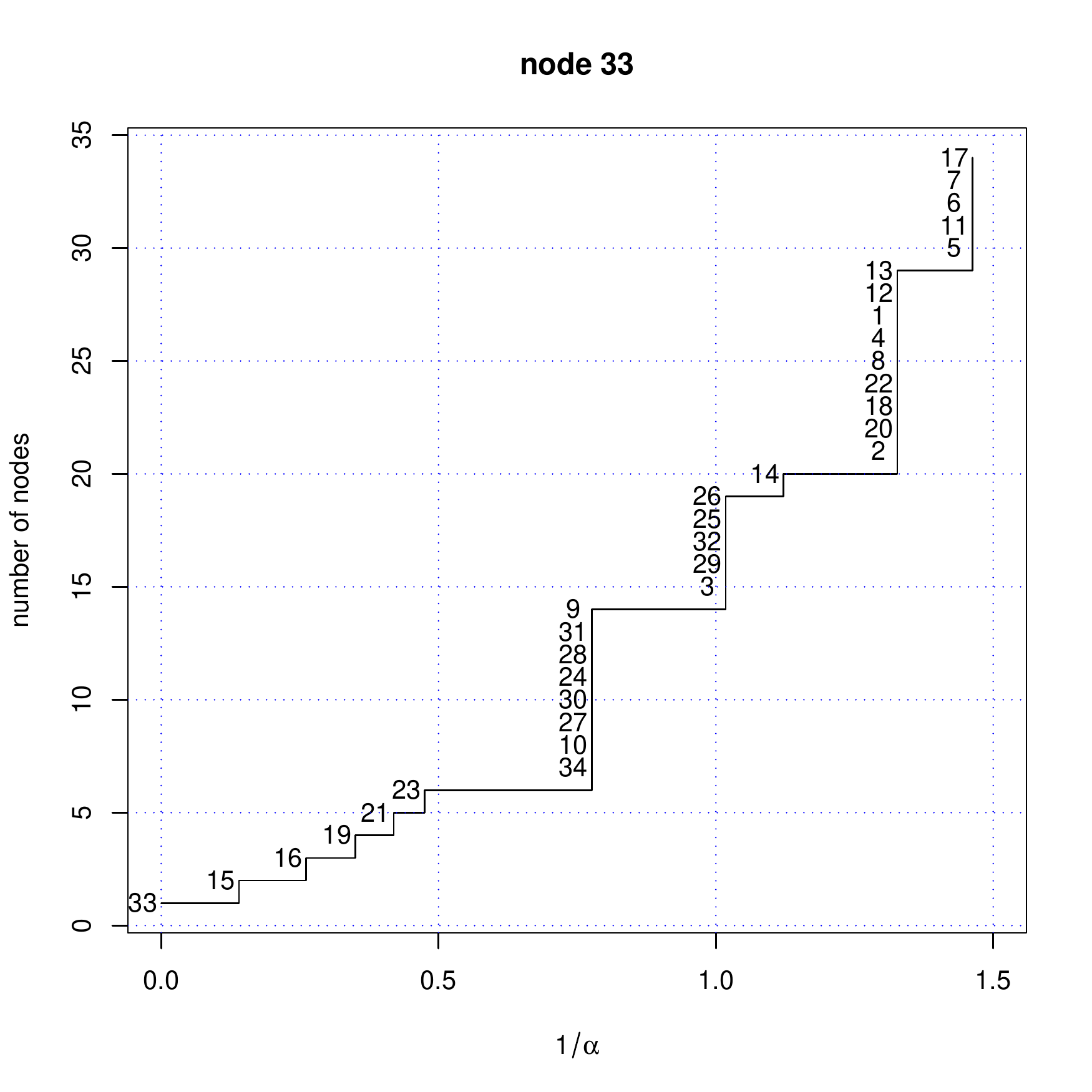}
\includegraphics[width=3.1in]{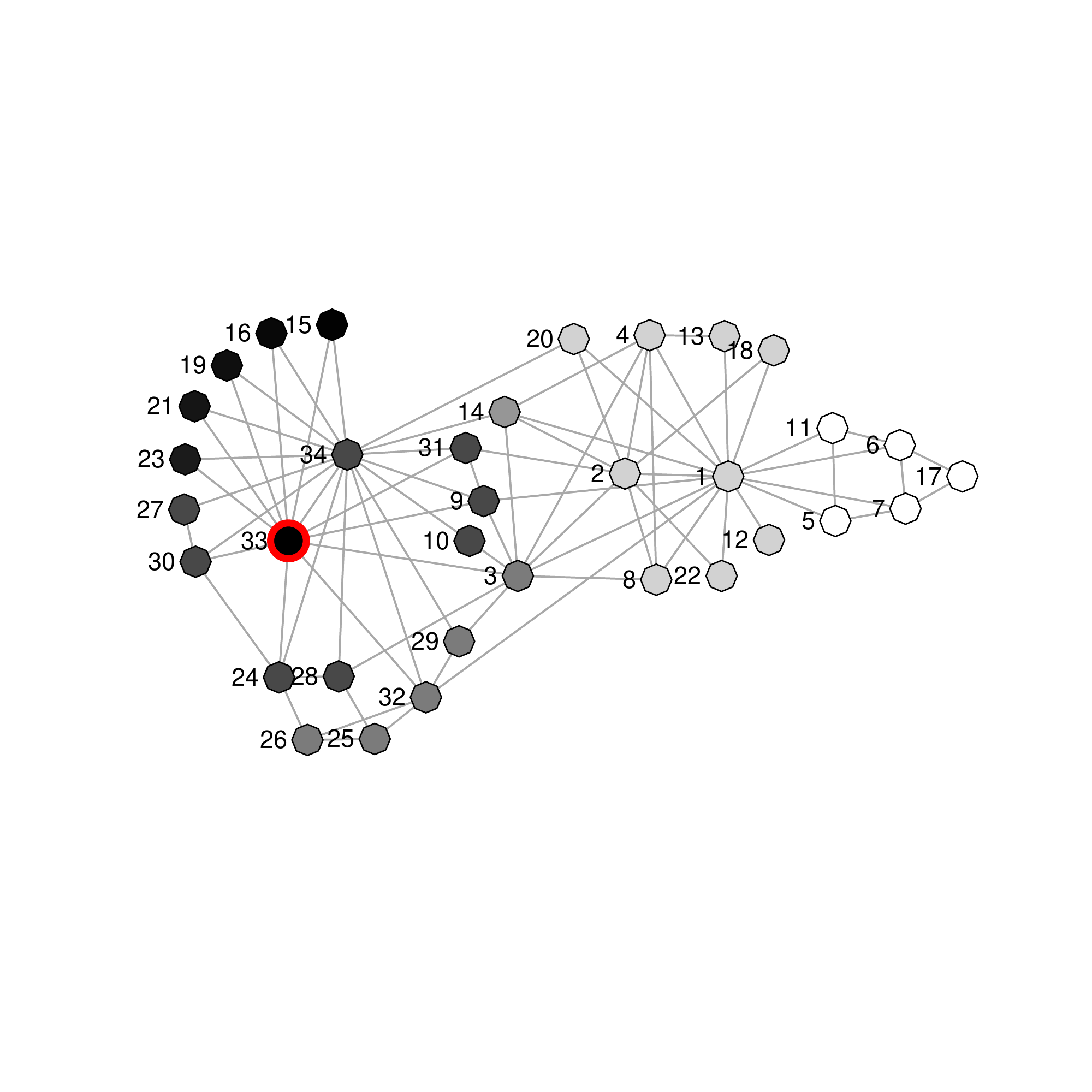}
\caption{Growing natural  MONC community of node 33 of karate club}
\label{fig_karate.comm.33}
\end{figure}

\begin{figure}[!t]
\centering
\includegraphics[width=3in]{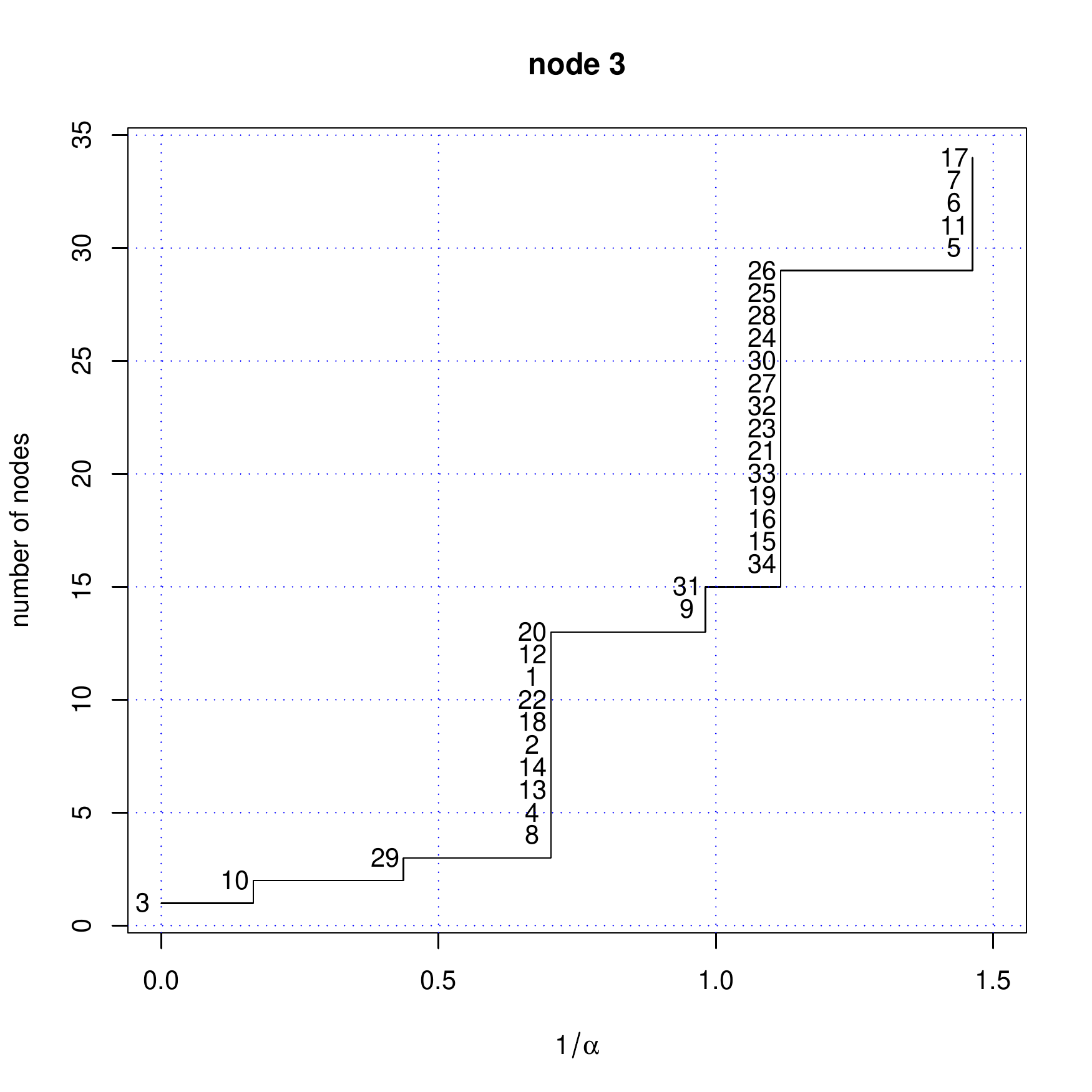}
\includegraphics[width=3.1in]{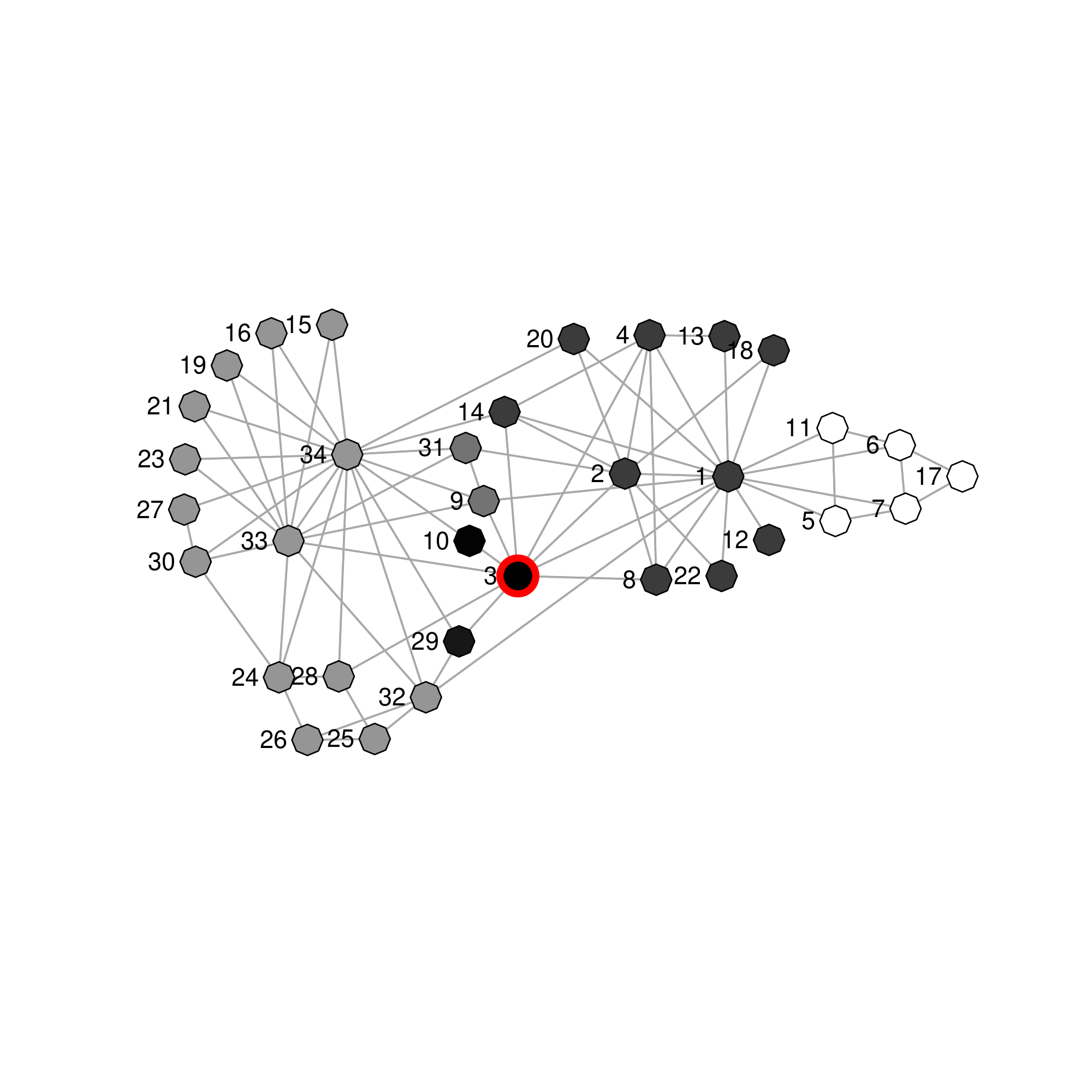}
\caption{Growing natural MONC  community of node 3 of karate club}
\label{fig_karate.comm.3}
\end{figure}

Since the network of 34 karate-club members is sparse---there is no clique with six or more fighters---we can apply MONC either by starting from each node or by using seed cliques. For the analysis starting with single seed nodes, figures \ref{fig_karate.comm.1}--\ref{fig_karate.comm.3} show the growing natural communities of three nodes. The step curve in the diagrams gives the growing number of nodes in the community as a function of $1/\alpha$. Each node is its own community at $1/\alpha = 0$. In our approach, the resolution always decreases, i.e.\ $1/\alpha$ cannot decrease.

For example (cf.\ figure \ref{fig_karate.comm.1}), even if nodes 11, 6, 7, and 17 enter the community of node 1 at lower $1/\alpha$ than their predecessor node 5, we display the same value of $1/\alpha$ for all five nodes because the higher resolution for the other four nodes becomes possible only after node 5 has been included. In other words, since the community's properties are changed in a way that would enable adding other nodes at a smaller value of $1/\alpha$ only as the result of adding node 5 at a higher value of $1/\alpha$, this higher value is the community-changing resolution level. 

The network graphs (figures \ref{fig_karate.comm.1}--\ref{fig_karate.comm.3}) visualise the growth of communities by displaying the seed node in black with a red corona, the last nodes joining in white, and the intermediate nodes on a grey scale corresponding to the resolution at which they come in. Lancichinetti \etal \cite[Fig. 6(a), p. 10]{lancichinetti2009detecting} display the LFM cover of the karate network they obtain in the resolution interval $.76 < \alpha < .84$ (which roughly equals the inverse resolution interval $1.2 < 1/\alpha < 1.3$). We see from the diagrams and graphs of nodes 1 and 33 that the MONC algorithm produces exactly the same cover in this interval, i.e.\ the same set of overlapping communities which cover the whole graph. 

Another cover in this resolution range, which is less frequently obtained by the LFM algorithm, becomes visible in the MONC diagram and graph of node 3 (figure \ref{fig_karate.comm.3}), a node in the overlap of the two communities of the cover displayed by Lancichinetti \textit{et al}. In this resolution range the community of node 3 contains all nodes except the five nodes on the right end of the karate graph. The communities of these five nodes are identical and contain no other node in the resolution interval considered.

These examples indicate that MONC gives approximately the same results as the LFM algorithm when applied to the karate-club network. A detailed comparison reveals that 31 of the modules we obtained with our implementation of LFM were also obtained by MONC. A Table in \textit{Supplementary Information} of reference \cite{havemann2010local} lists the corresponding resolution intervals for both algorithms. Small differences are partly due to the different fitness functions (s.\ \textit{Algorithms} section) and partly due to the randomness of LFM. Further 22 LFM modules were not produced by MONC. Their resolution intervals are mostly small~(maximum .2716, median .045). 
In addition MONC obtained 9 modules which LFM did not find. Each of these modules is found as an intermediate state of a growing natural community of only one seed node.

\begin{figure}[!t]
\centering
\includegraphics[width=4in]{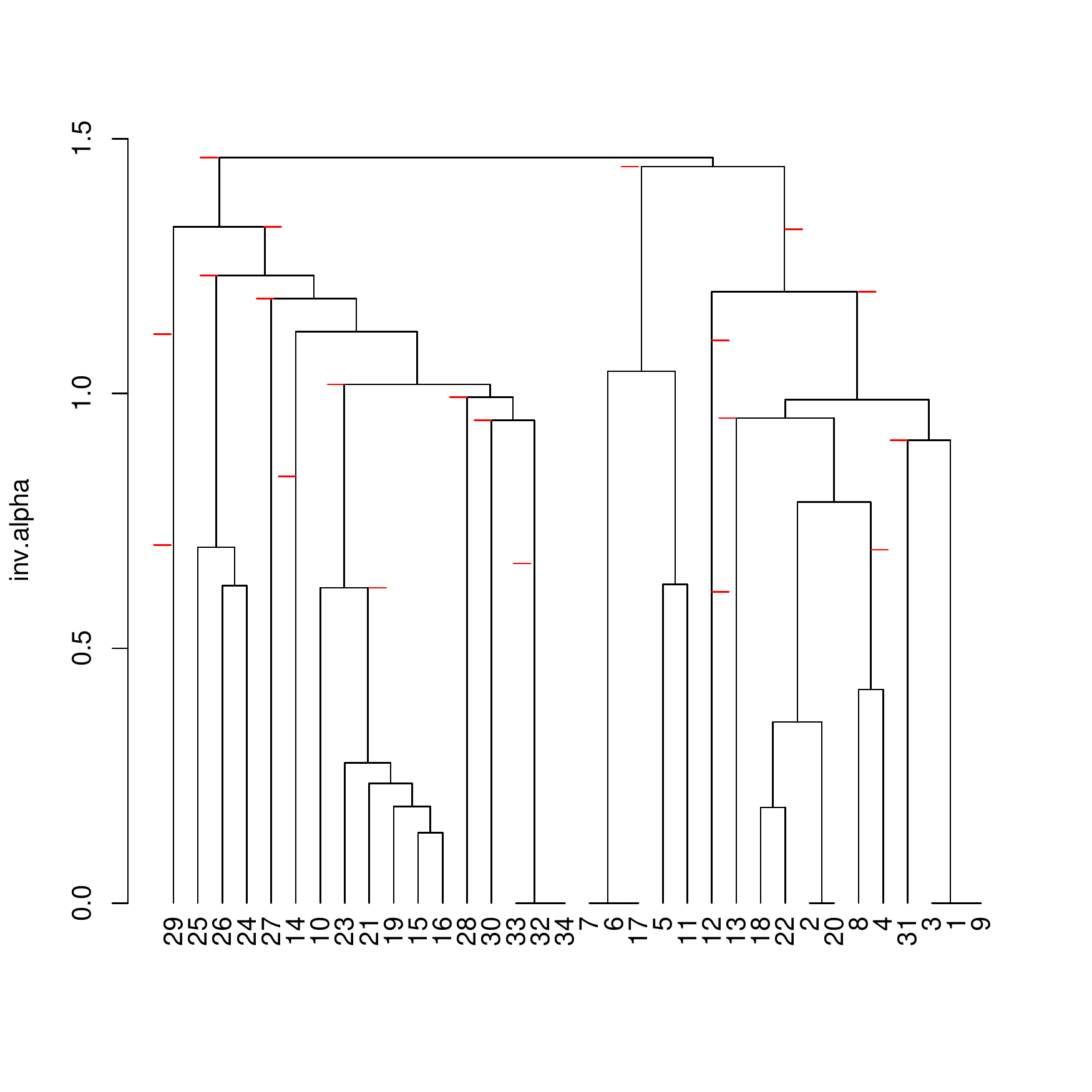}
\caption{Dendrogram of merging overlapping  MONC communities of the karate-club graph with cliques as seeds. Small red lines at its branches mark leaps where communities increase by at least three nodes.}
\label{fig_karate.dendro}
\end{figure}

We also tested whether cliques can be used as seeds for the karate-club graph. The results are very similar to those produced starting from single nodes as seeds, at least for higher levels of $1/\alpha$. Figure \ref{fig_karate.dendro} shows the dendrogram of the hierarchy of overlapping modules. The branches of the dendrogram represent growing natural communities of seed cliques (which may overlap one another and become identical at those values of $1/\alpha$ where the branches join) rather than disjunct sets of nodes. For each of the $n=34$ nodes one optimised seed clique is chosen, some nodes start with the same clique as e.g.\ nodes 1, 3, and~9~(cf.\ base line of figure \ref{fig_karate.dendro}). 

\begin{figure}[!b]
\centering
\includegraphics[width=3in]{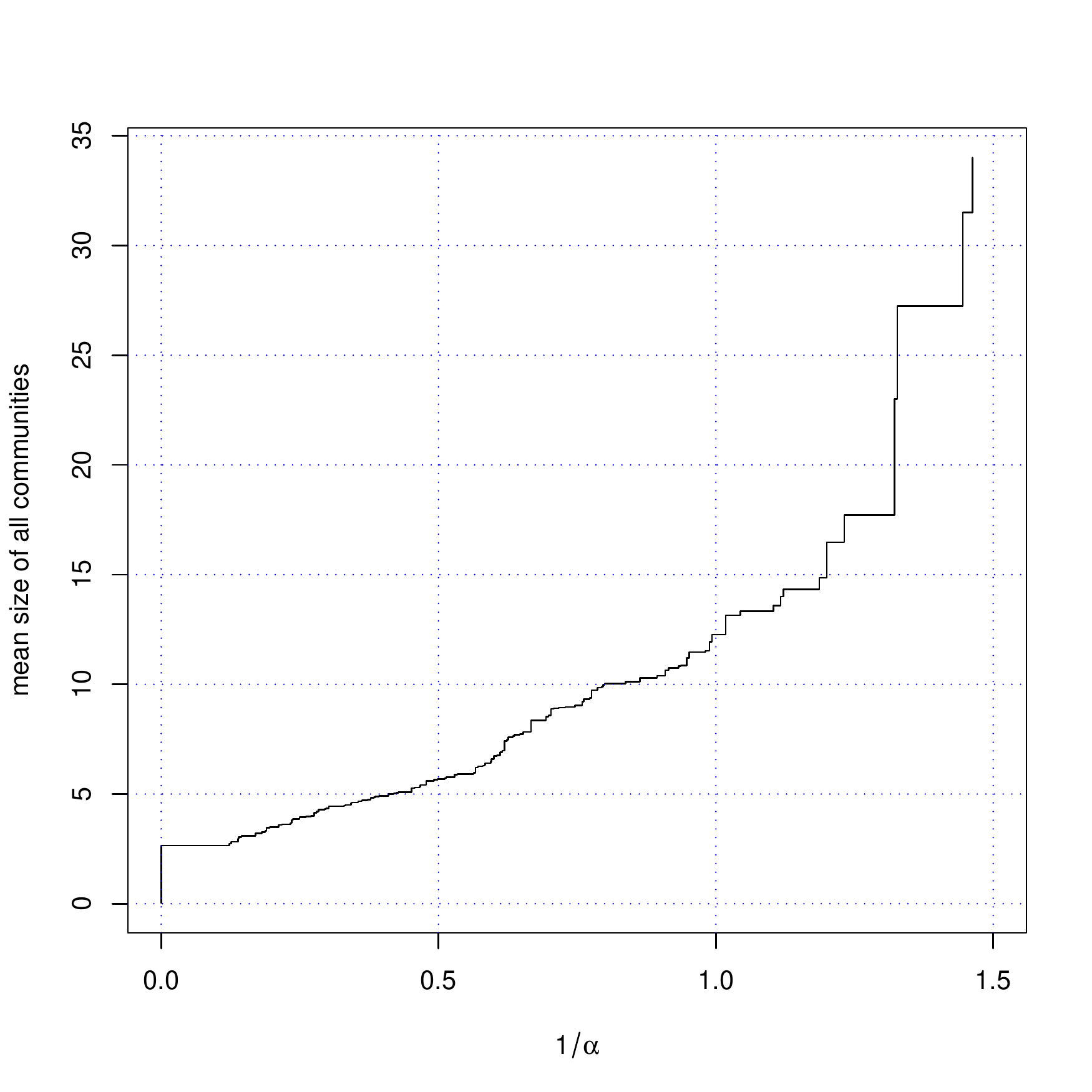}
\includegraphics[width=3in]{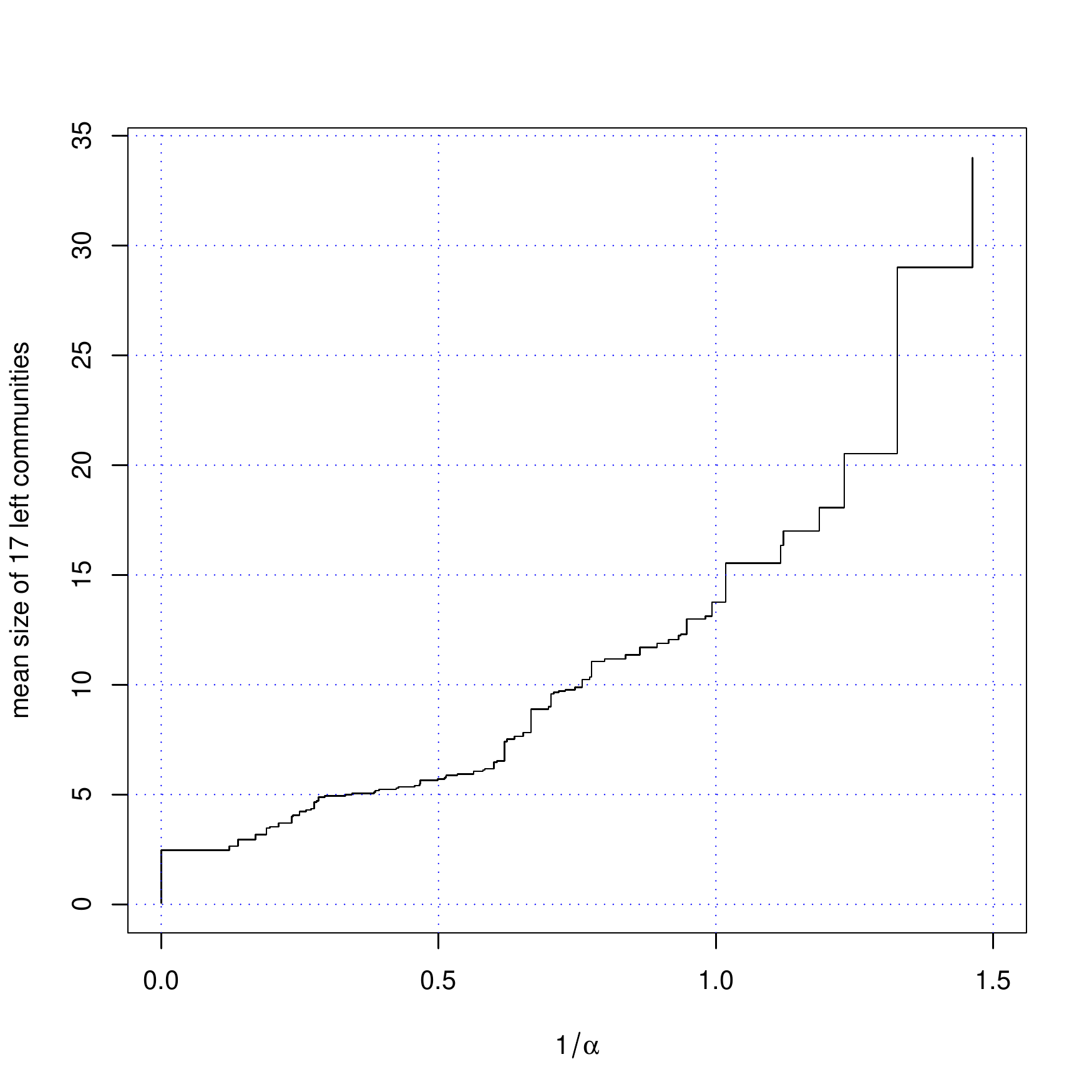}
\caption{Mean size of growing natural MONC communities in the karate-club graph over inverse resolution. Left: average of all 34 communities; right: average of 17 communities in the left branch of the dendrogram.}
\label{fig_karate.mean.size}
\end{figure}

Finding significant resolution levels of a graph is part of MONC's post-processing for which we tested several heuristic procedures. We propose the following method. MONC's $n$ growing natural communities can be displayed by step curves of community size over $1/\alpha$ (as in figures \ref{fig_karate.comm.1}--\ref{fig_karate.comm.3}). These curves are characterised by alternations of stable plateaus and rapid changes in size. Plateaus are particularly interesting because they indicate modules that remain stable over a broad resolution interval. This assumption is also made by Lancichinetti \etal \cite[p. 6]{lancichinetti2009detecting} and underlies one of the methods recently proposed by Lambiotte \cite[section IV]{lambiotte2010multi}. To apply this idea to whole covers of the graph, we have to unite the step curves of all $n$ communities. We do this by displaying the mean size of all growing communities over $1/\alpha$~(figure \ref{fig_karate.mean.size}). This step curve can be interpreted as the average `view' of $n$ different views on the graph. It shows big leaps and plateaus. The first leap at $1/\alpha = 0$ (followed by a plateau) is due to the seed cliques used here. Beyond that level, the broadest plateaus should mark relevant resolution intervals, which indeed they do.  

The broadest plateau begins at $1/\alpha = 1.327$ and ends at $1/\alpha = 1.445$. From the dendrogram (Figure \ref{fig_karate.dendro}) we learn that in this interval MONC detects three modules, namely: 
\begin{enumerate}
 \item the five nodes at the right end of the graph (5, 6, 7, 11, and 17; cf. figures \ref{fig_karate.comm.1}--\ref{fig_karate.comm.3}),
\item the whole rest of the graph (29 nodes), and
\item the whole graph.
\end{enumerate}
The first module is only `seen' by its own members, the second by the 17 nodes at the basis of the left branch of the dendrogram, and the whole graph by the remaining 12 nodes at the basis of the rightmost branch, which at this low resolution level do not `see' any separation in the graph. 

The next relevant plateau starts at $1/\alpha = 1.232$ and ends at $1/\alpha = 1.322$. In this interval, the dendrogram has four branches, i.e.\ four different modules. Two of them are identical to the first and the second module detected in the first interval but now the second module (the whole graph without the five nodes at the right end) is only `seen' by node 29 (the leftmost branch). The 16 nodes of the next branch (nodes 25 -- 34) now `see' a module with 20 nodes and the 12 nodes of the most right branch a module with 19 nodes. These two are the same modules that are detected by the LFM algorithm as the third most frequently found cover (in the interval  $1.19 < 1/\alpha < 1.32$, cf. the paper by Lancichinetti \etal \cite[fig. 6]{lancichinetti2009detecting}).

The graph of average sizes of all natural communities as displayed at the  left side of figure \ref{fig_karate.mean.size} can only show clear plateaus at higher levels of $1/\alpha$ because at lower levels the leaps of different branches of the dendrogram overlap~(the small red lines in figure \ref{fig_karate.dendro} mark all leaps where more than two nodes are added to the communities). In order to obtain lower relevant resolution levels we can restrict the set of nodes whose `views' are used for calculating the averages of size. As an example the right diagram  in figure \ref{fig_karate.mean.size} shows the diagram of mean community size of the left branch of the dendrogram based on 17 nodes. The mean sizes of 17 communities are of course different from the mean sizes of 34 communities. However, the left graph also shows big leaps and broad plateaus---with the difference that, as expected, averaging a smaller number of `views' provides a clearer picture. Thus, we propose to reveal the hierarchical structure of a graph by determining relevant cuts not only of the whole dendrogram but also of its main branches. Such a local procedure corresponds to the locality of MONC. This approach needs to be tested further.

\subsection{Information science}
The density variation across the graph of 492 information-science papers from 2008 requires starting the MONC algorithm with seed cliques. The 1812 maximal cliques of the 492 bibliographically coupled information-science papers differ strongly in size. There are many small maximal cliques and some large ones. These cliques were reduced by the method described above. Most papers belong to more than one reduced clique. Each paper is assigned to the clique where it has its maximum $\alpha_\mathrm{excl}$. This leads to the selection of
357 reduced cliques.  275 cliques belong only to one node.  Three papers do not belong to any reduced clique and are therefore used as single paper seeds. The distribution of clique sizes before and after reduction and further details are given  in \textit{Supplementary Information} section of reference \cite{havemann2010local}.

\begin{figure}[!b]
\centering
\includegraphics[width=3in]{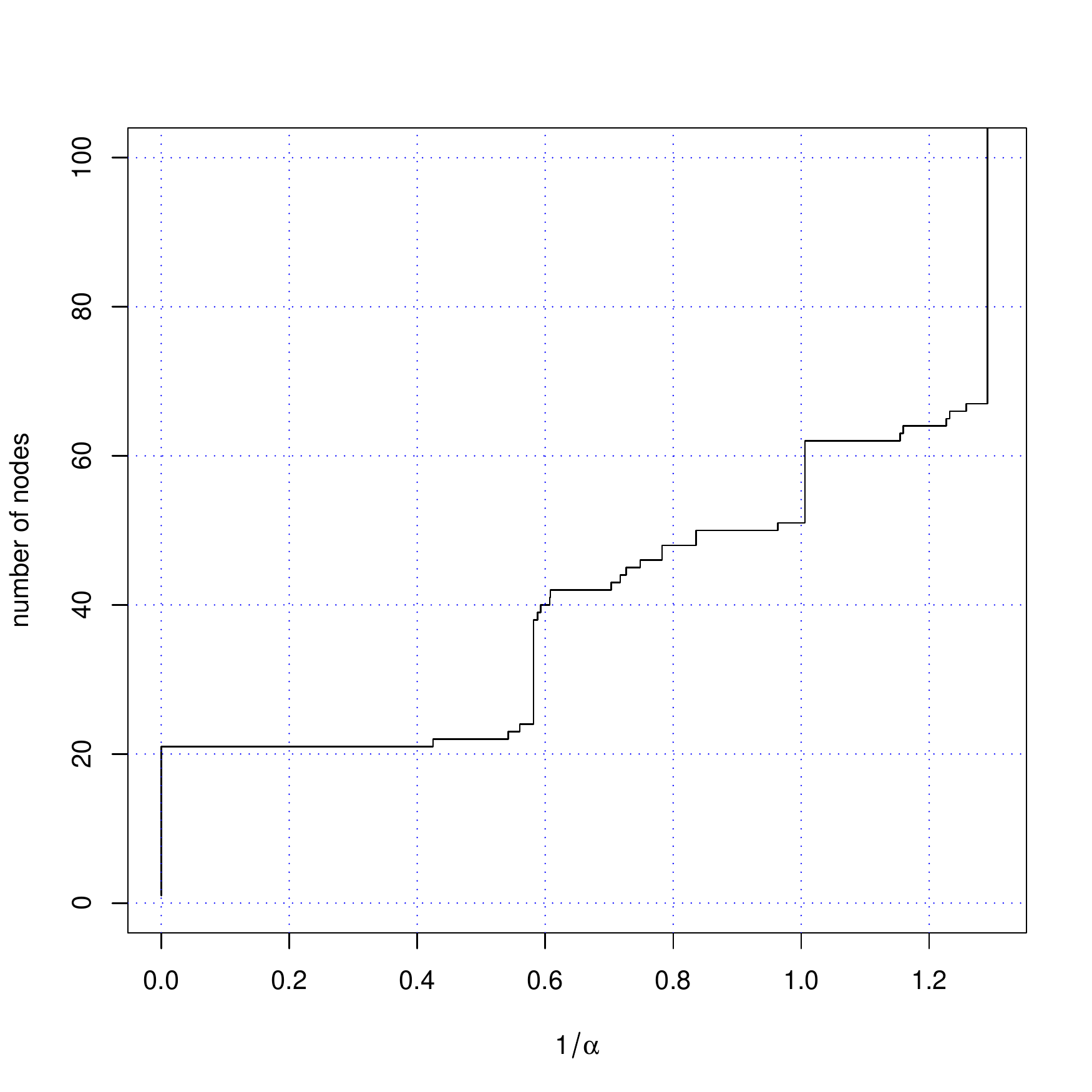}
\includegraphics[width=3in]{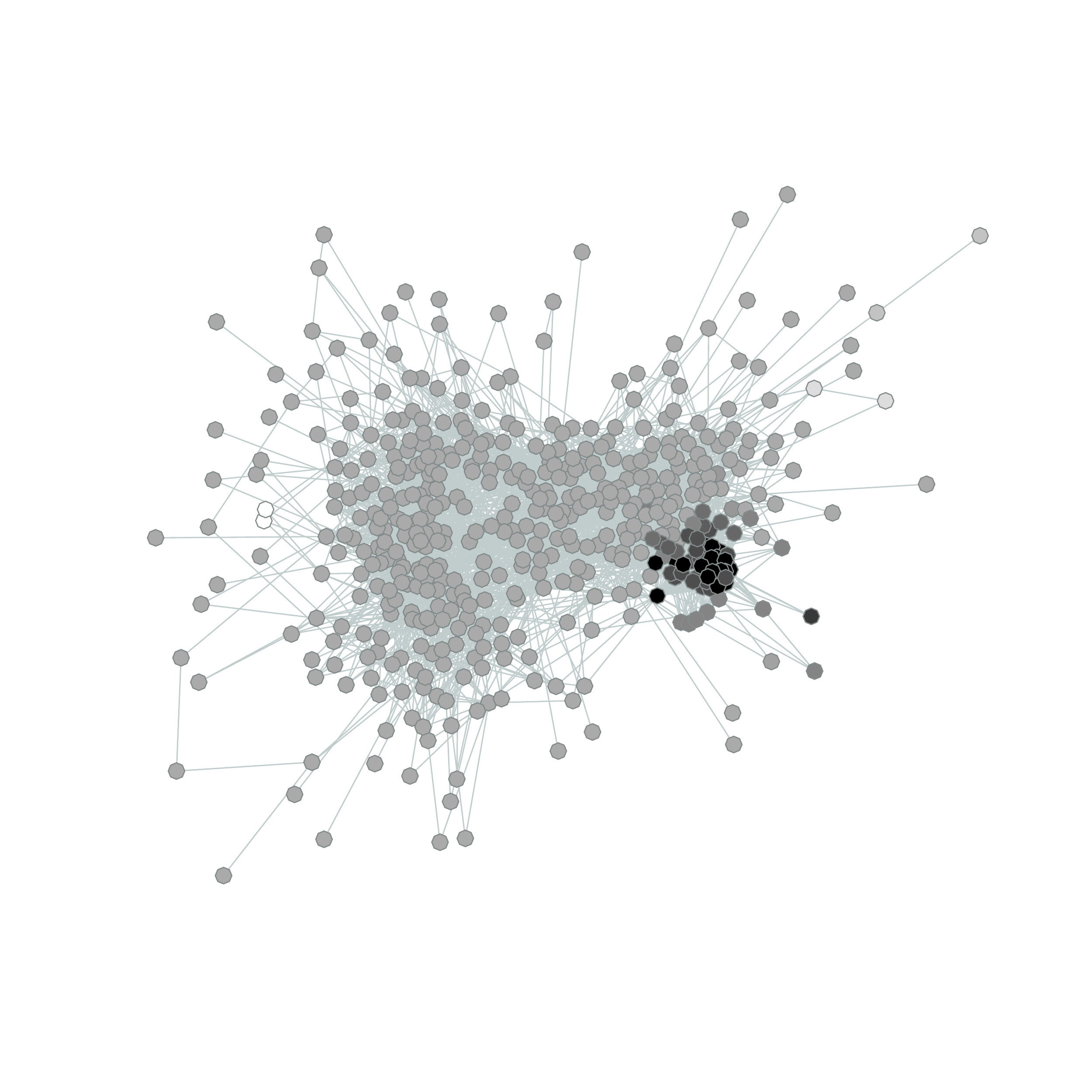}
\caption{Growing natural MONC community of \textit{h}-index clique (step curve up to 100 papers, network visualisation by force-directed placement)}
\label{fig_comm.21}
\end{figure}

\begin{figure}[!b]
\centering
\includegraphics[width=3in]{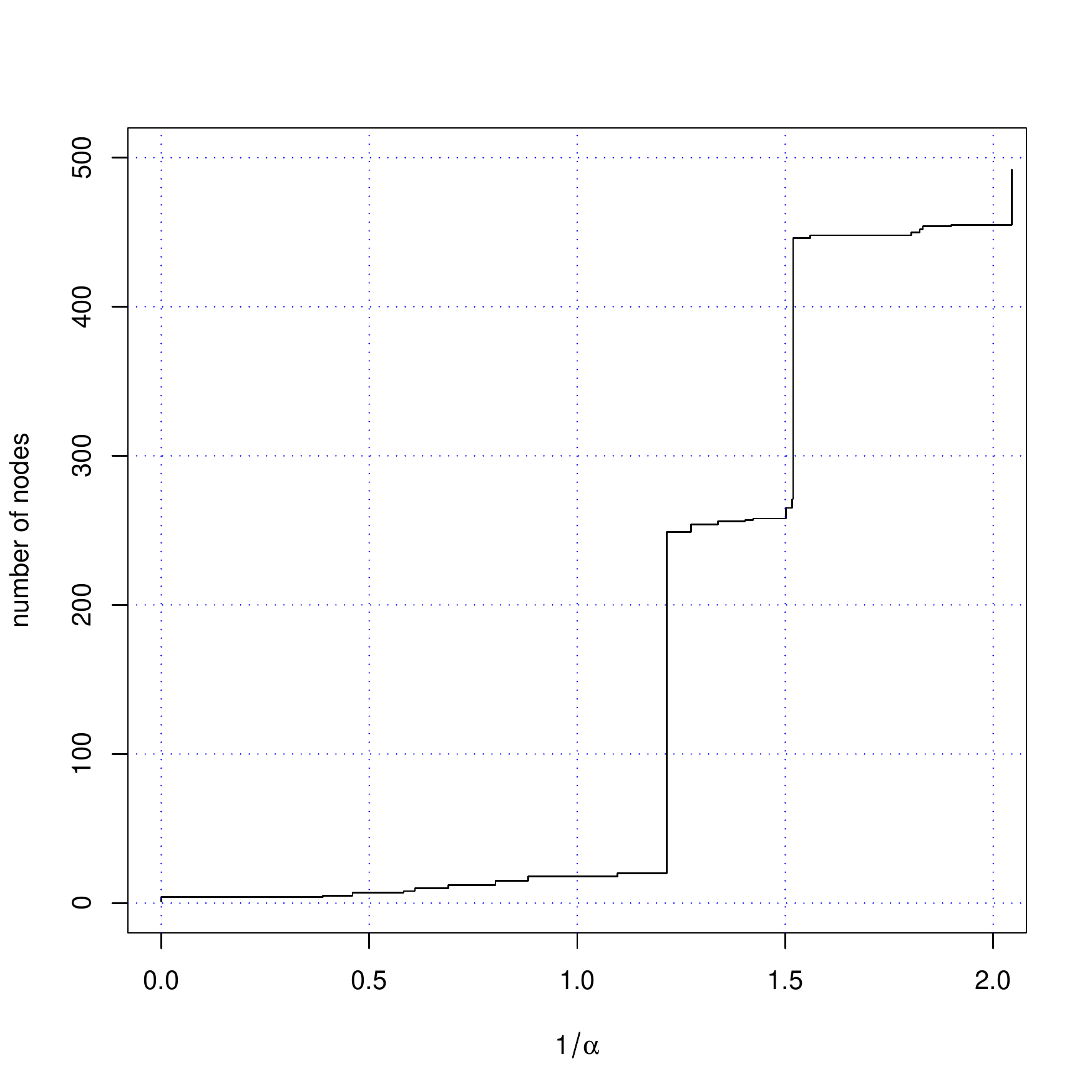}
\includegraphics[width=3in]{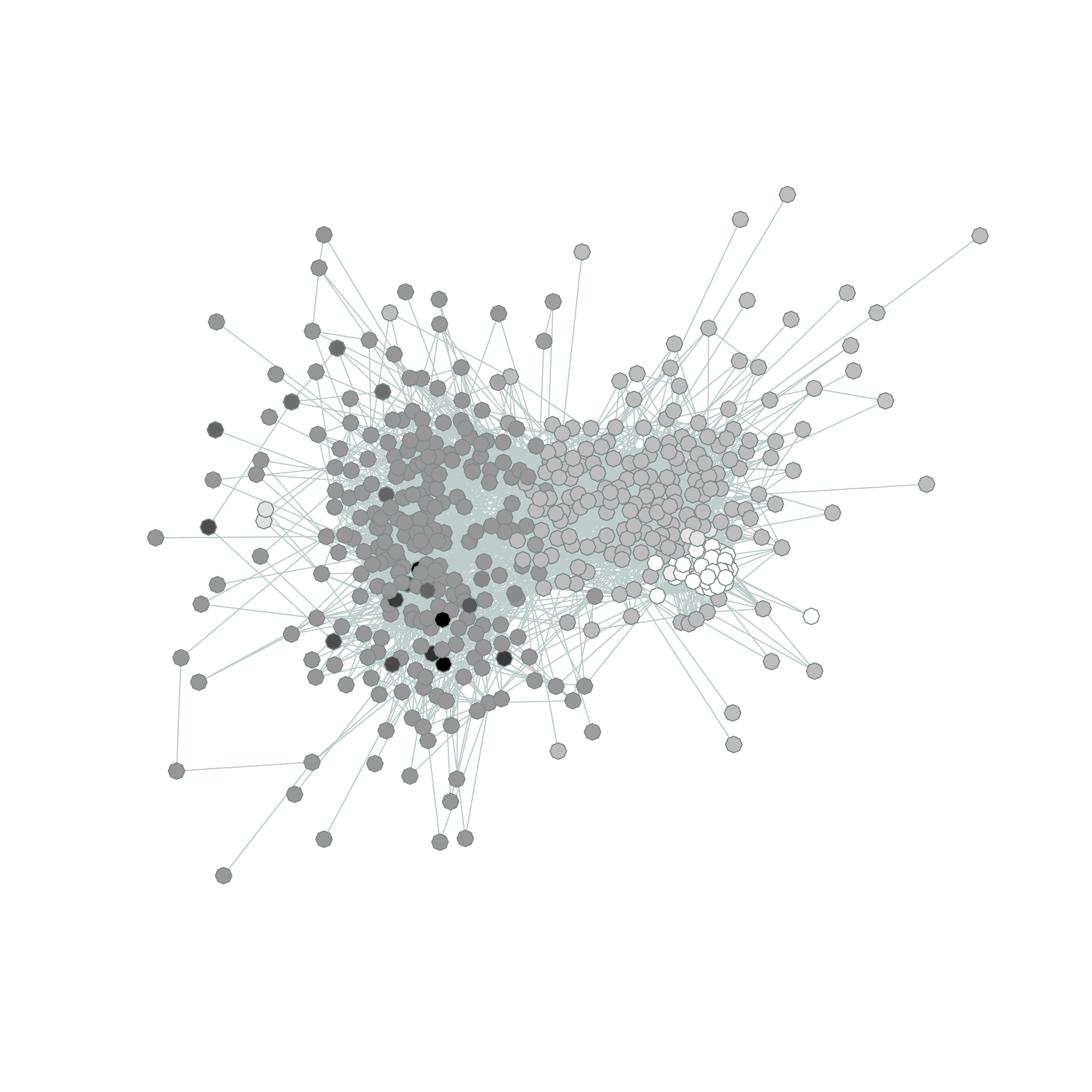}
\caption{Growing natural MONC community of IR-papers}
\label{fig_comm.1}
\end{figure}

The largest clique is formed by 46 papers which all cite the 2005 paper by J. E. Hirsch in which he proposes the \textit{h}-index. Many \textit{h}-clique papers also have the term \textit{h-index} in their titles but some other of them discuss it only as a method among others. We reduce the \textit{h}-clique by the method described above to 21 papers which all have the \textit{h}-index or its derivatives as a central topic (cf.\ figure \ref{fig_melting-h-clique} in \textit{Supplementary information}). 16 papers have their highest $\alpha_\mathrm{excl}$ in the \textit{h}-clique. We use the natural community of this clique as an example to demonstrate the results of MONC.

In Figure \ref{fig_comm.21} we show the graph of the growing natural community of one paper that has its highest $\alpha_\mathrm{excl}$ in the reduced \textit{h}-clique, whose 21 papers form the black core of the dark cloud in the figure. The corresponding diagram visualises the growing natural community up to 100 papers. After collecting further 21 papers more or less related to the topic (mostly citing the Hirsch paper) the community's growth decelerates. This slow development lasts till $1/\alpha \approx 1$ ending up with 51 papers. 

Figure \ref{fig_comm.1} shows a sequential graph displaying intermediate steps while growing a MONC community around a clique of information-retrieval (IR) papers. It shows the separation of IR papers (left) from papers in bibliometrics (right hand side). Note, that IR and bibliometrics are not separated if we start from $h$-clique. 

The LFM experiment started from $\alpha = 2$ and went down in steps of 1/100 to $\alpha = 0.1$. It produced the same succession of modules accumulating papers attached to the \textit{h}-community. The MONC algorithm proved indeed not only to be more accurate but also faster. A straightforward implementation of our MONC algorithm (without storing community parameters, see \textit{Algorithm} section) reduced computation time by about 75\,\% (both algorithms were implemented in \textbf{R}, cf. \url{http://141.20.126.172/~div}). In addition, the resolution thresholds computed by the MONC algorithm are much more accurate and the hierarchy of modules is detected automatically by MONC.

\begin{figure}[!t]
\centering
\includegraphics[width=2.6in]{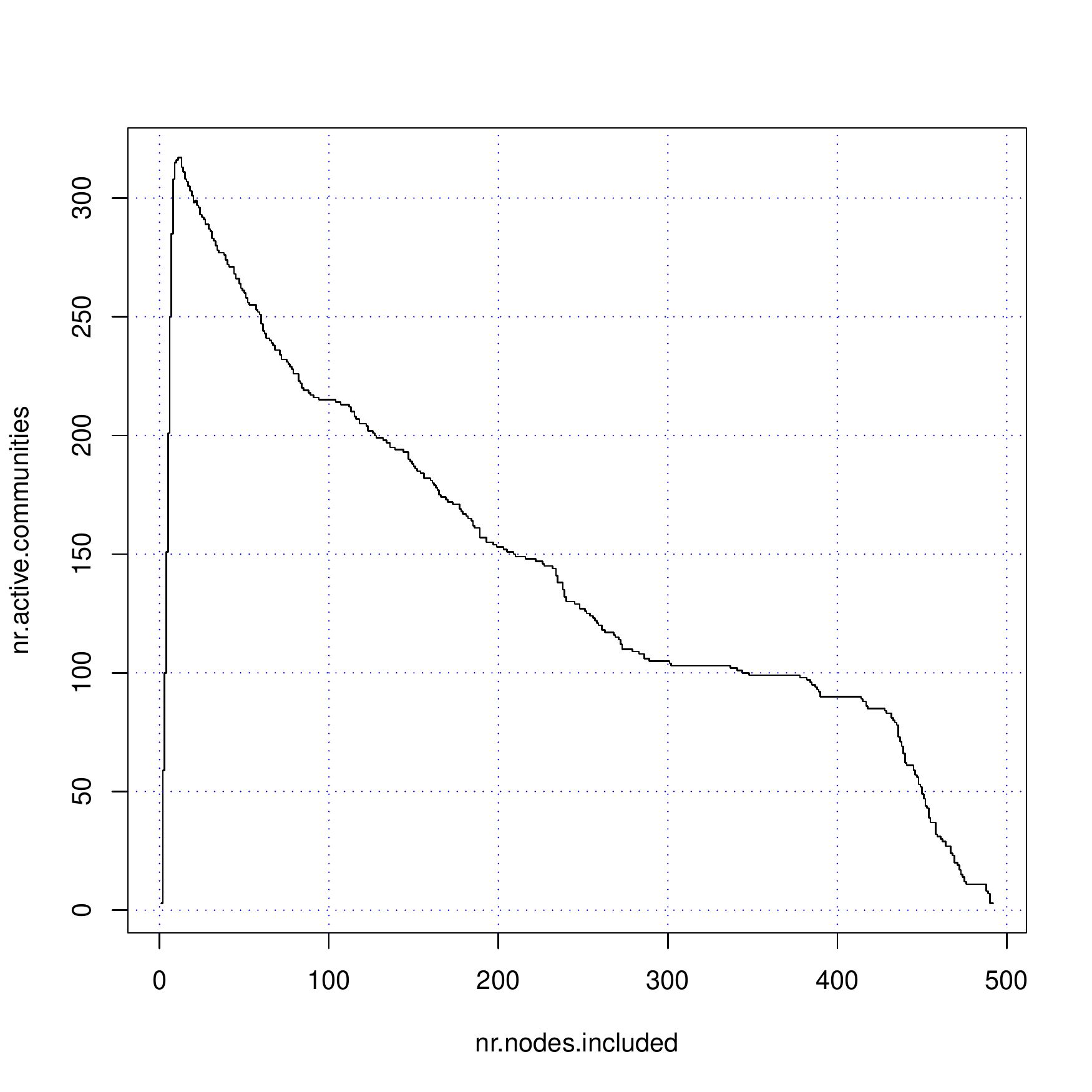}
\includegraphics[width=3.35in]{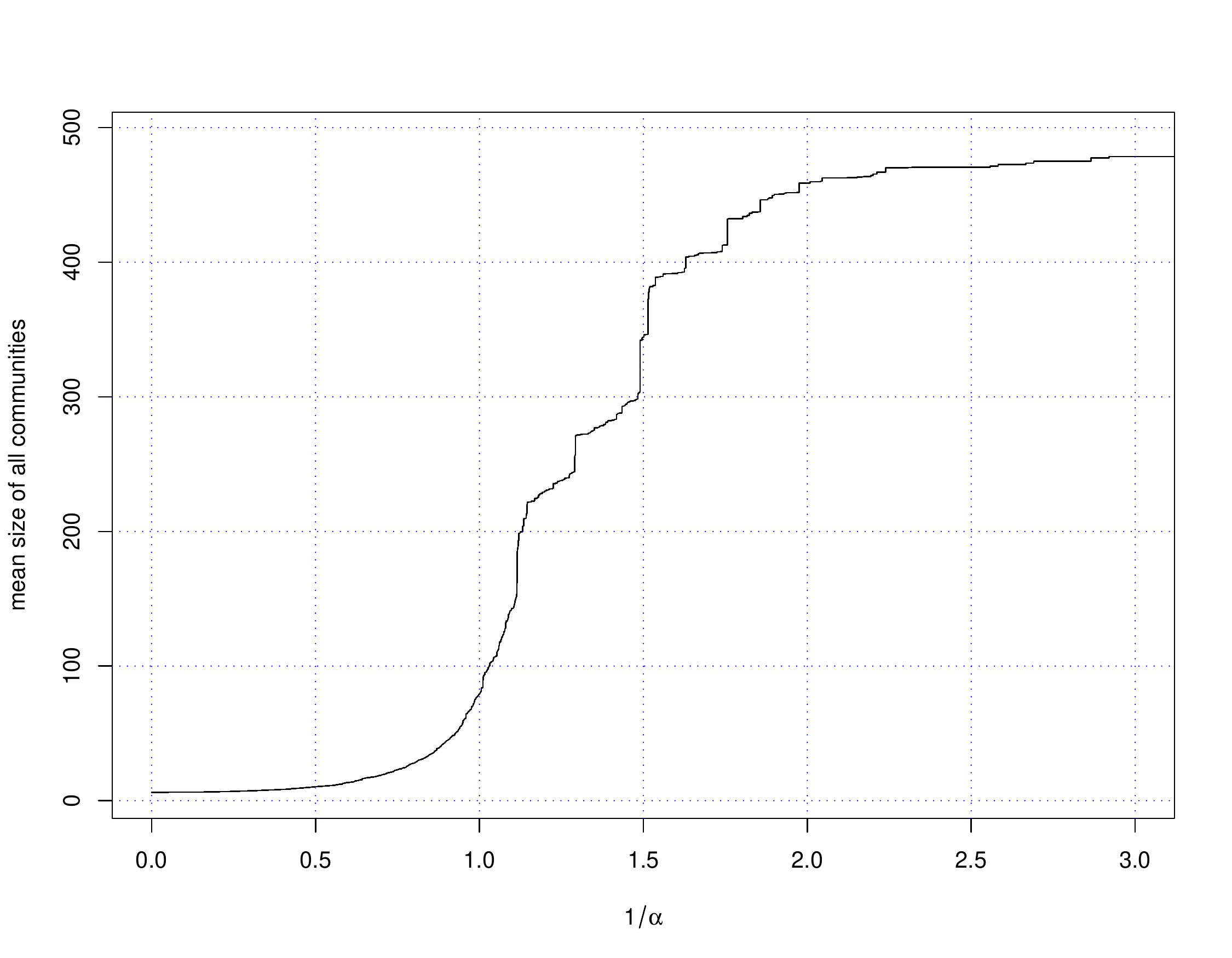}
\caption{Left side: Number of active MONC communities in information-science experiment as a function of nodes included. Right side: Mean size of all $n=492$ growing natural MONC communities over $1/\alpha$.}
\label{fig_nr.active.comm}
\end{figure}

To illustrate the merging of communities by MONC, the left diagram in  figure \ref{fig_nr.active.comm} displays the dynamics of the number of active communities, i.e.\ of communities which are still growing separately and up to the current number of nodes have not yet merged, which first rises to above 300~(of a maximum of 492 node communities or of 360 different seed cliques) and then falls rapidly, which again makes MONC run faster. By `active' we denote growing communities which up to the current number of nodes have not merged but will merge later. Only three communities survive before they are merged into the whole set of all 492 nodes.

The right diagram in figure \ref{fig_nr.active.comm} shows---as for the karate graph---the step curve of mean community sizes over inverse resolution $1/\alpha$. Again, we see big leaps but in contrast to the corresponding karate diagram the plateaus are less clear. We will test whether they become clearer if we determine mean sizes of communities separately for the branches of the MONC dendrogram of information science 2008. We will also test Lambiotte's robustness measure based on running means \cite[section IV]{lambiotte2010multi}.

\subsection{Synthetic benchmark graphs}
In order to further assess the performance of MONC, we tested its ability to recover the input modules of 1100 LFR benchmark graphs~(cf.\ \textit{Data} section). These benchmark graphs do not have an inbuilt hierarchical structure. Instead, they have only a trivial hierarchy: single nodes at high resolution, the whole graph at zero resolution, and a cover of overlapping modules at some resolution interval that has to be determined. The selection of adequate resolution levels has been discussed in the previous sections. We start with level $\alpha = 1$, at which the fitness of a module (eq.~\ref{LFK-fitness}) has to be larger then 1 to match the weak definition of communities given by Radicchi \etal \cite{Radicchi2004defining}, because we want to compare MONC's results with GCE's at its default resolution. 

With these benchmark graphs, one of MONC's major advantages---to reveal the whole hierarchical structure at all levels of resolution in one run---cannot be tested. Tests with hierarchical benchmark graphs are left to further work. However, testing MONC on non-hierarchical benchmark graphs provides many insights in the construction of definite modules at a certain resolution level, which is a necessary part of all evaluations of results of MONC. The experiments on real networks described in the preceding sections proved that MONC gives approximately the same results as LFM (karate club), and that its results are meaningful (information science). In neither of the experiments it was necessary to determine all modules at some fixed resolution level.

The main problem we have to solve when applying MONC to benchmark graphs is reducing the number of modules. For a selected resolution level, MONC constructs far more modules than the benchmarks have but many of them are near-duplicates. The same problem is reported for the GCE algorithm~\cite{lee2010detecting}. The production of many near-duplicates by MONC and GCE is probably due to the strict locality of both algorithms. The near-duplicates represent the slightly differing `perceptions' of the same `real' module by different seeds. In addition, we found modules which unite smaller modules `seen' by other seeds. We use the overlap measure of module distance Lee \etal~\cite[p.\ 3]{lee2010detecting} have found useful to identify near-duplicate and uniting modules:
\begin{equation}
\delta (G, G') = 1 - \frac{|G \cap G'|}{\min(|G|, |G'|)}.
\label{delta}
\end{equation}
It gives the proportion of nodes of the smaller module that are not elements of the larger module. Thus, modules of very different sizes can be identified. This is useful because some `real' separations of modules are not `seen' by all seeds, as was the case with the information-science graph, where bibliometrics and IR where not `seen' as separate by the  $h$-clique.

Instead of selecting one module from each set of modules identified using the overlap measure (Lee \etal~\cite[p.\ 4]{lee2010detecting}), we construct a \textit{consensus module} from each such set of modules. 
In \textit{Supplementary information} section we discuss the details of this procedure (s.\ p.\ \pageref{consensus-modules}). It extracts sets $C$ of near-duplicate modules with differing `views' on some assumed real module $M.$ In the case of fuzzy overlapping modules where nodes can have different degrees of membership~\cite{gregory2010fuzzy} we can use these differing views to estimate the degrees of membership of nodes in the real module $M.$ 
We give a node $V$ in the union of all modules of an extracted set $C$ a degree of membership $\mu$ in module $M$ equal to the proportion of $C$ modules which `select' node~$V.$

This definition of module membership is interesting because in most cases, degrees of memberships of a node in all finally determined modules do not sum up to 1. Normalisation is of course possible if a real network is more adequately described with normalised membership degrees. We leave tests of the fuzzy case to further work.

If the network is crisp by nature we need a threshold $\mu_c$ of membership for the determination of the real modules. In the case of the LFR benchmarks we need crisp  modules because the algorithms we want MONC to compare with produce crisp results.
We tested thresholds between $\mu_c=0.4$ and  $\mu_c=0.75$.

\begin{figure}[!t]
\centering
\includegraphics[width=4.5in]{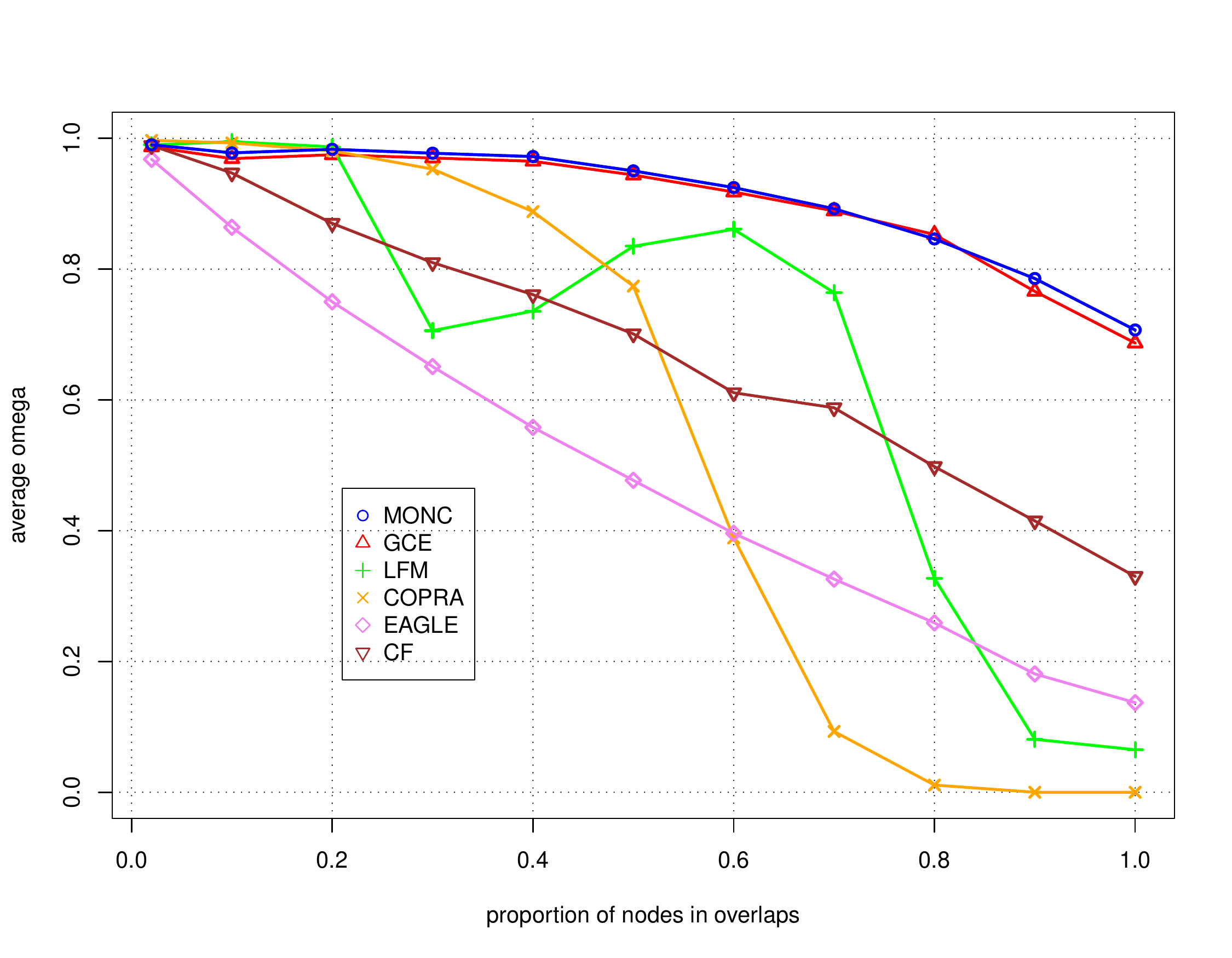}
\caption{Average omega index of covers of LFR graphs obtained by MONC and five other algorithms  which can detect overlapping modules. For each of the ten different degrees of overlap each algorithm runs on 100 randomly constructed graphs. Results of the five other algorithms supplied by Steve Gregory.}
\label{fig_omega-LFR-benchmarks}
\end{figure}

We compare the graph cover of overlapping modules determined by MONC at $\alpha=1$, $\delta_c= 0.25$, and $\mu_c=0.55$ with the corresponding benchmark input cover by using the omega index~\cite{collins1988omega}, which was also used by Steve Gregory in reference~\cite{gregory2010fuzzy}~(cf.\ \textit{Supplementary information} section, p.\ \pageref{Omega-index}). In figure~\ref{fig_omega-LFR-benchmarks} the average omega index of covers of samples of LFR benchmarks based on MONC results is displayed for different proportions of nodes in overlaps and compared with results of several other algorithms (which were calculated by Steve Gregory and are published here for the first time). Due to many similarities between GCE and MONC we expected similar omega curves of both algorithms but did not expect that both curves are essentially equal when regarding the standard deviations between 0.005 for the lowest overlap and 0.05 for the highest. 
Figure 9 shows that GCE and MONC (both at $\alpha = 1$) are the only algorithms that can reveal covers at high overlaps with a satisfactory level of accuracy. All other algorithms, i.e.\ LFM (random LFK, resolution $\alpha = 1$) \cite{lancichinetti2009detecting}, COPRA (with maximum number of communities  a vertex can belong to $\nu=5$) \cite{gregory2010finding}, CF (Clique Finder, with clique size $k = 4$) \cite{palla2005uoc}, and EAGLE \cite{shen2009detect} perform worse than GCE and MONC in these interesting regions of the diagram. 
We also found that MONC's results for a higher resolution ($1/\alpha = 0.95)$ become worse for low overlap but better for 80\,\% of nodes and more in overlaps.

\section{Discussion}
The algorithms compared here---LFK, LFM, GCE, and MONC---have in common that they construct growing natural communities by applying greedy procedures that are based on local fitness and are able to recover hierarchical structures of overlapping communities.

MONC and GCE are more strictly local procedures than LFM because they do not remove nodes from growing communities. This seems to be the reason why both MONC and GCE generally give better results with cliques as seeds, whereas LFM can start from single nodes. MONC is made faster by testing at each step of iteration whether two or more growing communities have already merged, i.e.\ have become equal. This merging can be visualised as a dendrogram of overlapping communities~(cf.\ figure \ref{fig_karate.dendro}, p.\ \pageref{fig_karate.dendro}). Thus, MONC can be seen as a truly hierarchical algorithm that clusters growing natural communities of a graph instead of its nodes.

In order to find modules of a network's nodes that can be interpreted, the user has to determine a set of resolution levels at which the hierarchical levels of the network shall be identified. Owing to the randomness of LFM its relevant levels can be obtained statistically by selecting the most frequent covers with their corresponding resolution intervals. For GCE, Lee \etal~\cite[p.\ 4]{lee2010detecting} propose to run the algorithm at different resolutions and to evaluate the modules by inspection.

For the results of MONC, we propose to determine relevant levels of resolution both globally for the whole graph and locally in different parts of the graph. Global levels correspond to cuts of the whole MONC dendrogram and local ones to separate cuts of its branches. The resolution levels are obtained by selecting the broadest plateaus in step curve diagrams of mean community size, with all communities for global levels and with communities of branches for local ones.

After choosing a set of relevant (global or local) resolution levels MONC's post-processing continues with determining all modules existing at each of these levels. We often find many near-duplicate modules at each level which are produced as near-equal spheres of different seeds and which will merge at some lower level of resolution. GCE faces the same problem and Lee \etal solve it by dismissing near-duplicates of modules during GCE's main processing~(when MONC only dismisses exact duplicates).

We interpret near-duplicate modules as different `views' on the same `true' community $M$ and use the set of near-duplicates which contain some particular node $V$ for constructing its degree of membership $\mu(V, M)$ with respect to $M$. Thereby we arrive at communities that are fuzzy sets over the universe of all nodes. Interestingly, memberships of a node in different communities do not sum up to unity~(but can of course be normalised).

What appears as a drawback of GCE---the existence of many near-duplicate modules---is an advantage of MONC because it can be used for detecting fuzzy overlapping modules in a network~(cf. Gregory's discussion of crisp and fuzzy overlaps in reference \cite{gregory2010fuzzy}). If the modules of the real network are not fuzzy MONC's results can easily be made crisp by choosing a critical threshold for membership~$\mu$.

Thus, although MONC itself is free of parameters for its post-processing we need to fix three parameters to obtain definite crisp modules which can be evaluated or interpreted: resolution $\alpha$, a threshold $\delta_c$ for near-duplicate distance, and a threshold $\mu_c$ for membership degree. Results for different thresholds $\delta_c$ and $\mu_c$ do not differ strongly in wide intervals for the 1100 LFR benchmarks.

We applied MONC to two real networks and 1100 synthetic benchmark graphs. For the real networks---karate club and 492 information-science papers---we obtained meaningful results that are very similar to the corresponding LFM results. In the tests on the synthetic benchmarks we found that MONC and GCE outperform the other four algorithms tested especially for graphs with a high proportion of nodes in overlaps.   

\section{Conclusions}
We propose a new algorithm, called MONC, that detects overlapping communities in a weighted undirected network and simultaneously reveals its hierarchy. MONC is local, deterministic and in its core free of parameters. Using a local fitness function it greedily expands natural communities of seeds until the whole (connected) graph is covered by the community of each seed. 

MONC obtains the hierarchy of communities not by numerically testing different resolution levels but analytically by calculating the next lower resolution level at which a natural community gets its next node. This analytic procedure is more exact and much faster then its numerical alternatives.

The strict locality and the exact calculation of community-changing resolution levels enable the application of MONC to a specific problem. In some empirical cases, the module structure of the whole network is less interesting than that of a subgraph with unknown borders which have to be determined first. In bibliometrics this is a typical situation when we want to analyse a paper network of a research specialty. Delineating this specialty in the global network contained in a bibliographic database is a persistent problem. Since the global network is too big, papers must be selected from the database before running a cluster algorithm, as we did in one of our experiments by selecting papers from six information-science journals. However, delineating specialties by journal sets is very imprecise and has only a low recall (s.\ already Bradford, 1934 \cite{Bradford1934sources}). MONC's locality allows to delineate subgraphs by expanding one relevant seed until its growing natural community remains relatively stable over a broad interval of resolution. This can be achieved by iteratively downloading the changing neighbourhood of the growing community from the database. Testing this procedure, which is likely to be useful for a whole class of applications, is one of the next points on our agenda.

\ack

The support of Steve Gregory, who run five algorithms on 1100 LFR benchmark graphs and calculated their performance data (for comparison with MONC results), is gratefully acknowledged. Steve also gave us valuable comments on two interim versions of this paper. We also thank Santo Fortunato and Conrad Lee for their comments on our algorithm; and an anonymous referee for useful advice and his request to test MONC on synthetic benchmark graphs. 

This work is part of a project in which we develop methods for measuring the diversity of research~(s. \url{http://141.20.126.172/~div}). The project is funded by the German Ministry for Education and Research (BMBF). We would like to thank all developers of \textbf{R}: \url{http://www.r-project.org}.

\appendix
\setcounter{section}{1}
\section*{Supplementary information}

\subsection*{Calculating resolution levels}
We derive the formula (eq.\ \ref{alpha_incl}, p.\ \pageref{alpha_incl}) for calculating the maximum value of $\alpha$, where a node $V$ does not diminish the fitness of a module $G$ when included in it. For $V$ in neighbourhood of $G$ we demand therefore

\begin{equation}
f(G \cup V, \alpha) > f(G, \alpha).
\end{equation} 
With definitions given in \textit{Algorithm} section we then have
\begin{equation}
\frac{k_\mathrm{in}(G \cup V) + 1}{k_\mathrm{tot}(G \cup V)^{\alpha}} > 
\frac{k_\mathrm{in}(G) + 1}{k_\mathrm{tot}(G)^{\alpha}}
\end{equation}
and therefore 
\begin{equation}
\frac{k_\mathrm{in}(G \cup V) + 1}{k_\mathrm{in}(G) + 1} > 
\left[ \frac{k_\mathrm{tot}(G \cup V)}{k_\mathrm{tot}(G)} \right] ^{\alpha}.
\end{equation}
We take logarithm on both sides of this equation and get
\begin{equation}
\log\frac{k_\mathrm{in}(G \cup V) + 1}{k_\mathrm{in}(G) + 1} >
\alpha \log\frac{k_\mathrm{tot}(G \cup V)}{k_\mathrm{tot}(G)}.
\end{equation}
That means, if $\alpha < \alpha_\mathrm{incl}$ with 
\begin{equation}
\alpha_\mathrm{incl} = \frac{\log({k_\mathrm{in}(G \cup V) + 1)}-\log({k_\mathrm{in}(G) + 1)}}{\log{k_\mathrm{tot}(G \cup V)}-\log{k_\mathrm{tot}(G)}}
\end{equation}
we have $f(G \cup V, \alpha) > f(G, \alpha)$. 

\subsection*{MONC pseudo code}
\label{pseudo-code}

After finding maximum cliques of all $n$ nodes of a (connected) graph, optimising them and selecting one optimised clique for each node as a pre-processing procedure of MONC the core algorithm starts. 

The growing natural communities of nodes are stored as an $n\times n$ matrix $C$ with one row for each community $i$, containing the $n$ node id-numbers in the succession of growth. In parallel the corresponding values of $1/\alpha_\mathrm{incl}$ are stored as matrix $\alpha^{-1}_{ij}.$ 

In the case of using nodes as seeds MONC's core can then described with the following pseudo code: 
\begin{algorithmic}[1]
\STATE {set all $n$ communities as active}
\STATE {set all $C_{i1} \gets i$}
\FOR {each column $j > 1$ of $C$}
 \FOR {each active community $G$}
    \FOR {each node $V$ adjacent to $G$} 
    \STATE {calculate $\alpha_\mathrm{incl}(G,  V)$} 
     \ENDFOR 
    \STATE {include the node with maximum $\alpha_\mathrm{incl}$ into $G$} 
   \ENDFOR
 \IF {any two active communities are equal }
 \STATE {make the community with larger $\alpha^{-1}_{ij}$ inactive}
 \ENDIF
\ENDFOR
\end{algorithmic} 

If cliques are used as seeds the id-numbers of their members have to filled in the rows of $C$ before starting the greedy algorithm. A community becomes active not until the iteration arrives at the first empty place in the corresponding row of $C$.

\begin{figure}[!b]
\centering
\includegraphics[width=2.5in]{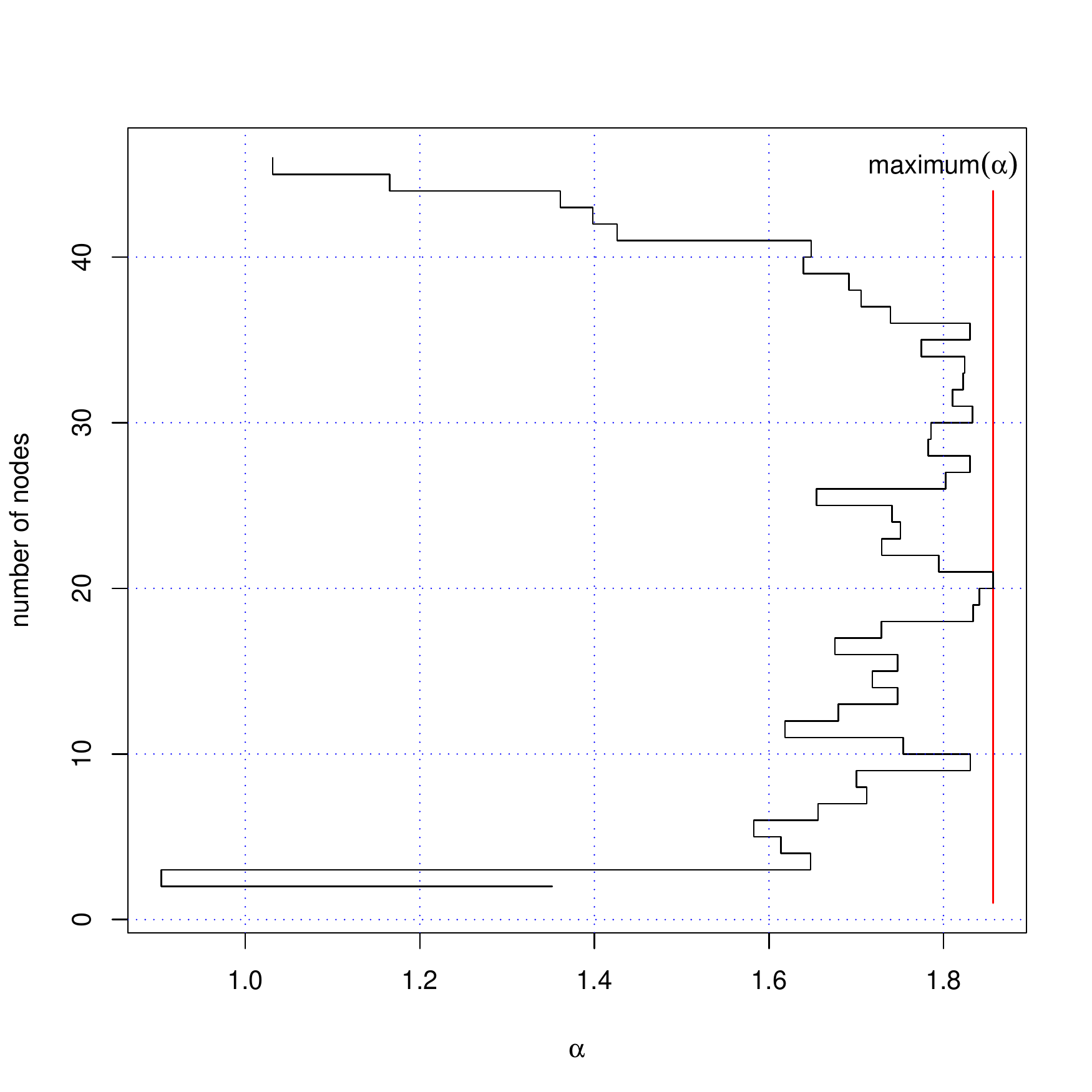}
\caption{Optimisation of a clique of 46 $h$-index papers to 21 core papers.}
\label{fig_melting-h-clique}
\end{figure}

To get a reduced clique from a maximal one we use an algorithm described by the following pseudo code:
\begin{algorithmic}[1]
\WHILE {clique size $> 2$}
 \FOR {each member $V$ in clique $c$}
    \STATE {calculate and store $\alpha_\mathrm{excl}(c,  V)$} 
    \STATE {exclude the node with minimum $\alpha_\mathrm{excl}$ from clique $c$} 
   \ENDFOR
 \ENDWHILE
\STATE {fill clique $c$ with nodes which have been excluded before $\max(\alpha_\mathrm{excl}(c,  V))$ was reached}
\end{algorithmic} 
The procedure is illustrated by figure  \ref{fig_melting-h-clique}.

\subsection*{Finding consensus modules} 
\label{consensus-modules}
Using a threshold for module overlap $\delta$ we first determine a similarity graph $S$ of all modules found by MONC at some resolution level. We have tested different thresholds between $\delta_c = 0.2$ and $\delta_c = 0.3$. 

For some benchmarks the similarity graph $S$ has as many components as the corresponding benchmark has modules and many $S$ components are cliques of near-duplicates. But often cliques of near-duplicate modules are linked because the overlap distance measure $\delta$ (eq.\ \ref{delta}) exceeds $\delta_c$ also in cases where one large module is compared to two small modules in different cliques of near-duplicates. In other words, a large module can overlap very different smaller modules belonging to different $S$ cliques. We remove those large bridge modules from similarity graph $S$ with smaller neighbours that are not direct neighbours. They emerge from seeds which cannot `see' the `real' separation of modules. Analogously we could also omit small bridges from $S$ but did not test this yet. 

After deleting large bridge modules we consider all components of similarity graph $S$ as sets of near-duplicate modules which represent often only slightly differing views on `real' modules. 
 
\subsection*{Information-science journals}

\begin{table}[!h]
\caption{\label{table.journals}533 Papers (528 articles and 5 letters) in volume 2008 of six information-science journals (source: Web of Science).}
\begin{indented}
\item[]\begin{tabular}{@{}rr}
\br
journal                 &                papers  \\
\mr
\textit{Information Processing \& Management}     &    111 \\   
\textit{Journal of Documentation}	         &    40 \\ 
\textit{Journal of Information Science	}         &    49 \\ 
\textit{Journal of Informetrics}                  &    31 \\ 
\textit{Journal of the American Society for   
 Information Science and Technology }     &   176  \\ 
\textit{Scientometrics	}                         &   126 \\ 
\mr
sum                                      &   533 \\ 
\br
\end{tabular}
\end{indented}
\end{table}

\subsection*{Omega index} 
\label{Omega-index}
This measure of similarity between two graph covers $C_1$ and $C_2$ with overlapping modules is a straight-forward generalisation of the well-known adjusted Rand index \cite{hubert1985comparing} used for partitions of disjoint modules~(cf.\ reference \cite{gregory2010fuzzy}). The unadjusted Rand index can be written as
\begin{equation}
r_u(C_1, C_2) = \frac{1}{N} \Bigl\{ |t_0(C_1) \cap t_0(C_2)|+|t_1(C_1) \cap t_2(C_2)| \Bigr\},  
\end{equation}
where $N = n(n-1)/2$ is the number of all possible pairs of the $n$ nodes of the graph and $t_j(C)$ is the set of node pairs which occur $j$ times together  in a module in partition $C$ ($j = 0, 1$). Because for overlapping modules node pairs can occur in more than one module, for the unadjusted omega index the sum has to run over all possible values till the maximum number $m_{max}=\max (m(C_1), m(C_2))$ of modules existing in $C_1 $ or $C_2$:
\begin{equation}
\omega_u(C_1, C_2) = \frac{1}{N} \sum_{j=0}^{m_{max}} |t_j(C_1) \cap t_j(C_2)|.  
\end{equation}
Adjusting the omega index means subtraction of the expected  omega index
\begin{equation}
\omega_e(C_1, C_2) = \frac{1}{N^2} \sum_{j=0}^{m_{max}} |t_j(C_1)|\cdot |t_j(C_2)|  
\end{equation} 
and normalising it by the maximum of this difference:
\begin{equation}
\omega(C_1, C_2) = \frac{\omega_u(C_1, C_2)-\omega_e(C_1, C_2)}{1-\omega_e(C_1, C_2)}.  
\end{equation} 


\section*{References}

\bibliography{informetrics}
\end{document}